\documentclass[conference]{IEEEtran}

\newcommand{\subparagraph}{}

\IEEEoverridecommandlockouts
\usepackage{cite}
\usepackage{amsmath,amssymb,amsfonts}
\usepackage{algorithmic}
\usepackage{graphicx}
\usepackage{textcomp}
\usepackage{xcolor}

\def\BibTeX{{\rm B\kern-.05em{\sc i\kern-.025em b}\kern-.08em
    T\kern-.1667em\lower.7ex\hbox{E}\kern-.125emX}}

\usepackage{tikz}
\usepackage{amsmath}
\usepackage{tablefootnote}

\usepackage{longtable}

\usepackage{filecontents}

\newcommand{\eat}[1]{}
\newcommand{\sys}{\textit{Silicon Dating}}
\definecolor{purple}{rgb}{1,0,1}

\newcommand{\ie}{i.e.,}
\newcommand{\eg}{e.g.,}

\usepackage{blindtext, graphicx, xcolor, listings, booktabs, multirow, colortbl, bm}
\usepackage{algorithm, algorithmic}
\usepackage{arydshln}
\usepackage{subfig}
\usepackage{array}
\usepackage{caption}
\usepackage{mdframed, ntheorem}
\usepackage{amssymb}% http://ctan.org/pkg/amssymb
\usepackage{pifont}% http://ctan.org/pkg/pifont
\usepackage{placeins}
\usepackage{diagbox}
\usepackage{gensymb}
\usepackage{url}

\usepackage{titlesec}
\titlespacing\section{0pt}{0pt plus 2pt minus 2pt}{0pt plus 0pt minus 2pt}
\titlespacing\subsection{0pt}{0pt plus 2pt minus 2pt}{0pt plus 2pt minus 2pt}
\titlespacing\subsubsection{0pt}{0pt plus 2pt minus 0pt}{0pt plus 2pt minus 2pt}

\begin{document}
% \hypersetup{draft} % Temporary solution to "\pdfendlink ended up in different nesting level than \pdfstartlink" error
%-------------------------------------------------------------------------------

\title{Silicon Dating}
%don't want date printed
% \date{}

\eat{
\title{Conference Paper Title\\
\thanks{Identify applicable funding agency here. If none, delete this.}
}
}

\author{\IEEEauthorblockN{Harrison Williams}
\IEEEauthorblockA{\textit{Virginia Tech}\\
hrwill@vt.edu}
\and
\IEEEauthorblockN{Alexander Lind}
\IEEEauthorblockA{\textit{Virginia Tech}\\
alexlind@vt.edu}
\and
\IEEEauthorblockN{Kishankumar Parikh}
\IEEEauthorblockA{\textit{Virginia Tech}\\
kjp2376@vt.edu}
\and
\IEEEauthorblockN{Matthew Hicks}
\IEEEauthorblockA{\textit{Virginia Tech}\\
mdhicks2@vt.edu}
}

\maketitle

\pagestyle{plain}

% Matt's notes: do not delete
% chip -> device
% 0 before 1
% device before software
% burn in -> imprinting...maybe burn-in is better

\begin{abstract}
In order to service an ever-growing base of legacy electronics, both government and industry customers must turn to third-party brokers for components in short supply or discontinued by the original manufacturer.
Sourcing equipment from a third party creates an opportunity for unscrupulous gray market suppliers to insert counterfeit devices: failed, knock-off, or otherwise inferior to the original product.
This increases the supplier's profits at the expense of reduced performance/reliability of the customer's system.
The most challenging class of counterfeit devices to detect is recycled counterfeits: recovered genuine devices which are re-sold as new.
Such devices are difficult to detect because they typically pass performance and parametric tests but fail prematurely due to age-related wear.

To address the challenge of detecting recycled devices pre-deployment, we develop \sys{}: a low-overhead classifier for detecting recycled integrated circuits using Static Random-Access Memory (SRAM) power-on states.
\sys{} targets devices with no known-new record or purpose-built anti-recycling hardware.
We observe that over time, software running on a device imprints its unique data patterns into SRAM through analog-domain changes; we measure the level and direction of this change through SRAM power-on state statistics.
In contrast to highly symmetric power-on states produced by variation during SRAM fabrication, we show that embedded software data is generally highly asymmetric and that the degree of power-on state asymmetry imprinted by software reveals device use.
Using empirical results from embedded benchmarks running on several microcontrollers, we show that \sys{} identifies recycled devices with 84.1\% accuracy with no software-specific knowledge and with 92.0\% accuracy by incorporating software knowledge---without prior device enrollment or modification.
\end{abstract}

\eat{\begin{IEEEkeywords}
Counterfeit, Static Random-Access Memory (SRAM), Aging
\end{IEEEkeywords}
}

\section{Introduction}
% As volume and performance demands for semiconductor devices increase, the supply chain has expanded globally---a single Integrated Circuit (IC) can undergo each fabrication, packaging, and testing stage under the control of a different entity in a different part of the world before reaching a user.
Many electronics customers rely on third-party brokers instead of Original Equipment Manufacturers (OEMs) to supply components for a lower cost or shorter lead time.
Sectors such as defense or aerospace, which require supporting legacy equipment expected to run for decades, are often forced to depend on these "gray market" suppliers to source components discontinued or under-produced by the OEM~\cite{cyber-supply-chain, counterfeit-inquiry}.
Third-party suppliers add a weak link to the supply chain between manufacturer and customer: unscrupulous brokers selling directly to the customer or further up in the supply chain have an opportunity to increase profit margins by supplying inferior devices and marking them as OEM-quality.
Such counterfeit devices are a growing threat to the electronics industry: one 2013 report estimated the cost of counterfeit electronics to U.S.-based companies alone to be \$7.5 billion per year~\cite{sia-anti-counterfeiting}.

Of the wide variety of counterfeit devices, including failed, up-marked, and knock-off products, recycled components---old but genuine devices salvaged from circuit boards and re-sold as new---make up 80 to 90\% of all counterfeits currently in circulation~\cite{faked-parts-detection}.
Instead of out-of-spec operation, recycled components threaten systems with early failure by working long enough to pass functional testing and failing due to age shortly after deployment.
Depending on the end-application for the component, premature failure can have expensive or fatal consequences: known incidents of counterfeit devices in safety-critical applications include automated medical equipment and braking systems in high-speed trains~\cite{sia-anti-counterfeiting}.
These systems are composed of multiple smaller discrete components such as microcontrollers, sensors, and other application-specific elements.
Determining IC age informs the end-user's decision about the age of a larger system: fine-grain counterfeit detection at the IC level propagates up to the system level.

% Past work addresses the challenge of recycled IC detection through several strategies.
The growing impact of recycled electronics has given rise to several detection strategies.
One set of approaches targets future systems by designing in a ``silicon odometer'', a circuit that responds predictably to aging effects and reveals a device's operating time~\cite{fingerprint-sensor, on-chip-structures, fpga-aging-monitor, fpga-aging-analysis, fpga-characterization}.
Hardware approaches work when implemented by the OEM, but leave past and presently-manufactured devices unserved; the vast majority of ICs built today do not include aging-detection circuitry.
To avoid the design changes and die space overhead associated with hardware solutions, enrollment-based approaches combine an aging-sensitive metric with an aging-insensitive one, which are both recorded at manufacture-time; suspect ICs are identified by the age-insensitive metric, and a significant deviation from the original age-sensitive measurement indicates the device is recycled~\cite{sram-enrollment-2, scare}.
However, enrollment techniques are not available for older devices and require manufacturers to maintain a device ID database.
Other methods exist that avoid designed-in hardware structures or pre-deployment enrollment, but are limited in scalability and practicality  because they are destructive, time-intensive, and expensive~\cite{counterfeit-integrated-circuits, statistical-methods}.

To enable high-confidence, non-destructive, recycled IC detection without aging sensors or enrollment, we provide \sys{}.
\sys{} leverages five observations:
\begin{enumerate}
    \item Static Random-Access Memory (SRAM) underlies all computing systems that we use: it exists on microcontrollers, Field-Programmable Gate Arrays, and desktop-class processors.
    \item SRAM's tightly-cross-coupled construction filters-out systematic variation sources like wafer-scale variation and constant operational variation (e.g., ambient temperature). This means that while individual SRAM power-on states are non-deterministic (due to manufacturing- and run-time chaos), SRAM memories as a whole look similar (in a statistical sense) across devices.
    \item SRAM's power-on state statistics (e.g., the proportion of times a cell powers-on to $1$), provides a digital window into the analog-domain properties of a SRAM cell.
    \item The analog-domain properties of an SRAM cell change in a software-dependent manner as the device operates.
    \item The values used by software are often asymmetric. In bulk, we show a trend of 0-biased values across our software benchmarks. In detail, we show that there exist software-specific patterns where most SRAM cells hold the same value throughout the lifetime of the program.
\end{enumerate}
Putting these observations together: \textbf{over time, software asymmetries gradually burn-in to the SRAM, becoming asymmetries in its analog-domain properties.
This analog-level change is measurable in the digital domain through SRAM power-on state statistics.}
As a program runs on the device, SRAM power-on states trend away from the expected new distribution towards the biased data patterns inherent in software.
By measuring the degree of bias present in SRAM's power-on state, \sys{} estimates the age of the system.
To address cases where the auditor has partial knowledge of the device under test and possible software (e.g., popular legacy single-purpose systems), we demonstrate how to incorporate such knowledge into our program-unaware classifier to create an increased accuracy program-aware version of \sys{}.

We evaluate both the program-aware and program-agnostic versions of \sys{} using a suite of five benchmarks representative of common embedded software applications and run them under accelerated-aging conditions on the Texas Instruments MSP430G2553~\cite{g2553datasheet}.
Our evaluation shows that the program-agnostic version correctly classifies 72.9\% of 5-month old devices and 84.1\% of 5-year old devices, while the program-aware version increases accuracy to 82.8\% and 92.0\%, respectively.

This paper makes the following technical contributions.
\begin{itemize}
    \item We explore SRAM cell aging mechanisms and show that typical software applications exercise these aging mechanisms in consistent, predictable ways (\ref{sec:new_sram}).
    \item We design a software system to determine the age of currently deployed and future systems based on common data biases that software imprints on SRAM in the form of cell power-on state biases (\ref{sec:unaware-design}).
    \item We extend our design with analysis of unique hardware and software biases for the device under test; in this case, we examine the specific aging behavior of the system to produce a fine-grain estimate of device age (\ref{sec:structural-awareness}, \ref{sec:aware-design}).
    \item We implement this system on commodity devices and show that it: (1) reliably detects up to 92.0\% of recycled devices (\ref{sec:unaware-eval}), and (2) is robust to a variety of statistical and physical noise sources (\ref{sec:operating-time}, \ref{sec:baseline-variance}).
    \item We evaluate the effects of natural and adversarial SRAM recovery and show that reducing \sys{}'s accuracy by less than eight percent places a large time burden on the attacker (\ref{sec:natural_recovery}, \ref{sec:adversarial_recovery}).
\end{itemize}

\section{Background}
% \subsection{SRAM Cell Aging}
% \input{Figures/SRAM_Cell.tex}
% \input{Figures/preage_heatmap.tex}
\begin{figure}[t]
  \centering
  \subfloat[\footnotesize A six-transistor SRAM cell.\label{fig:sram_cell}]{%
    \includegraphics[width=0.485\columnwidth]{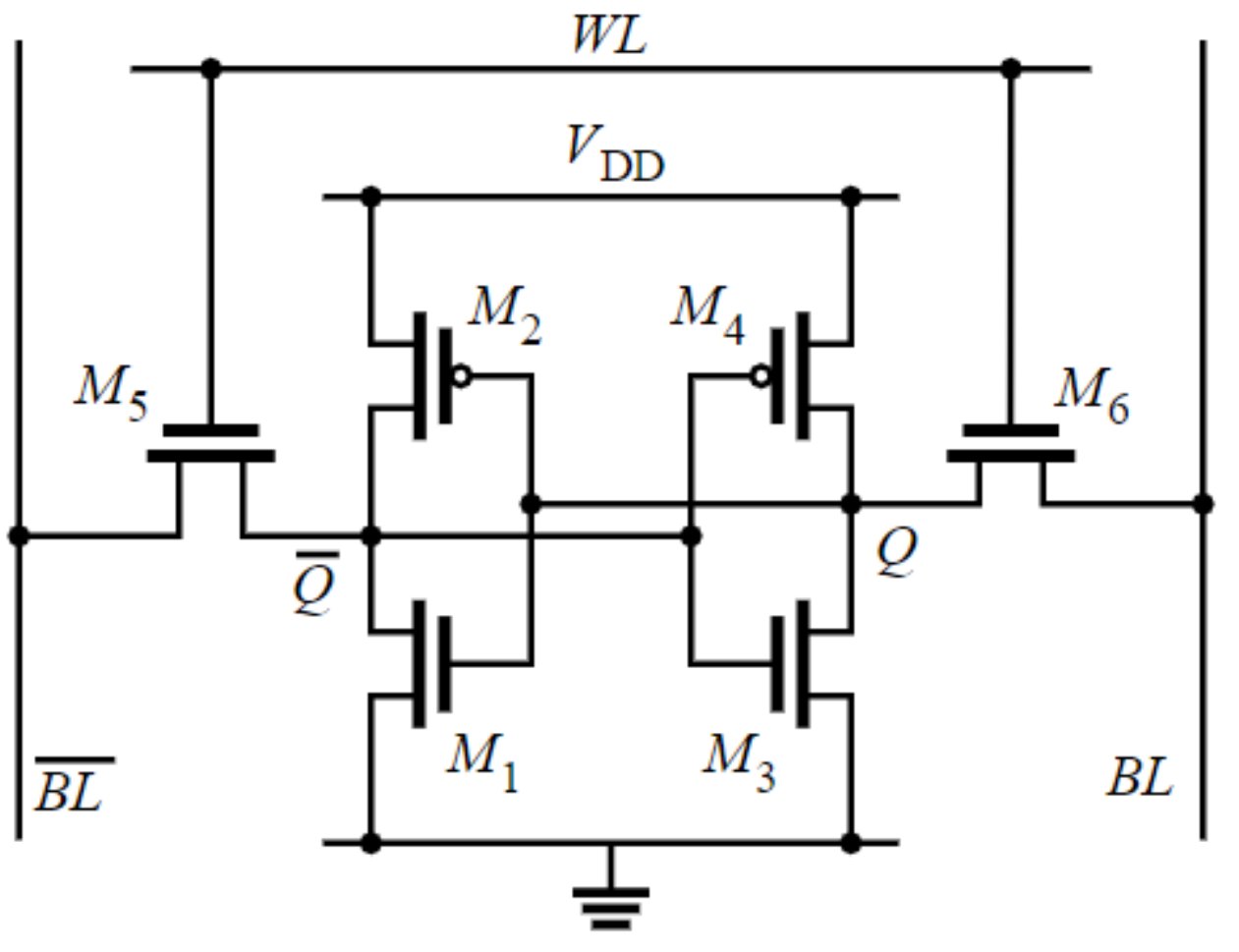}
  }
  \subfloat[\footnotesize MSP430G2553 SRAM power-on state before aging.\label{fig:preage_heatmap}]{%
    \includegraphics[width=0.485\columnwidth]{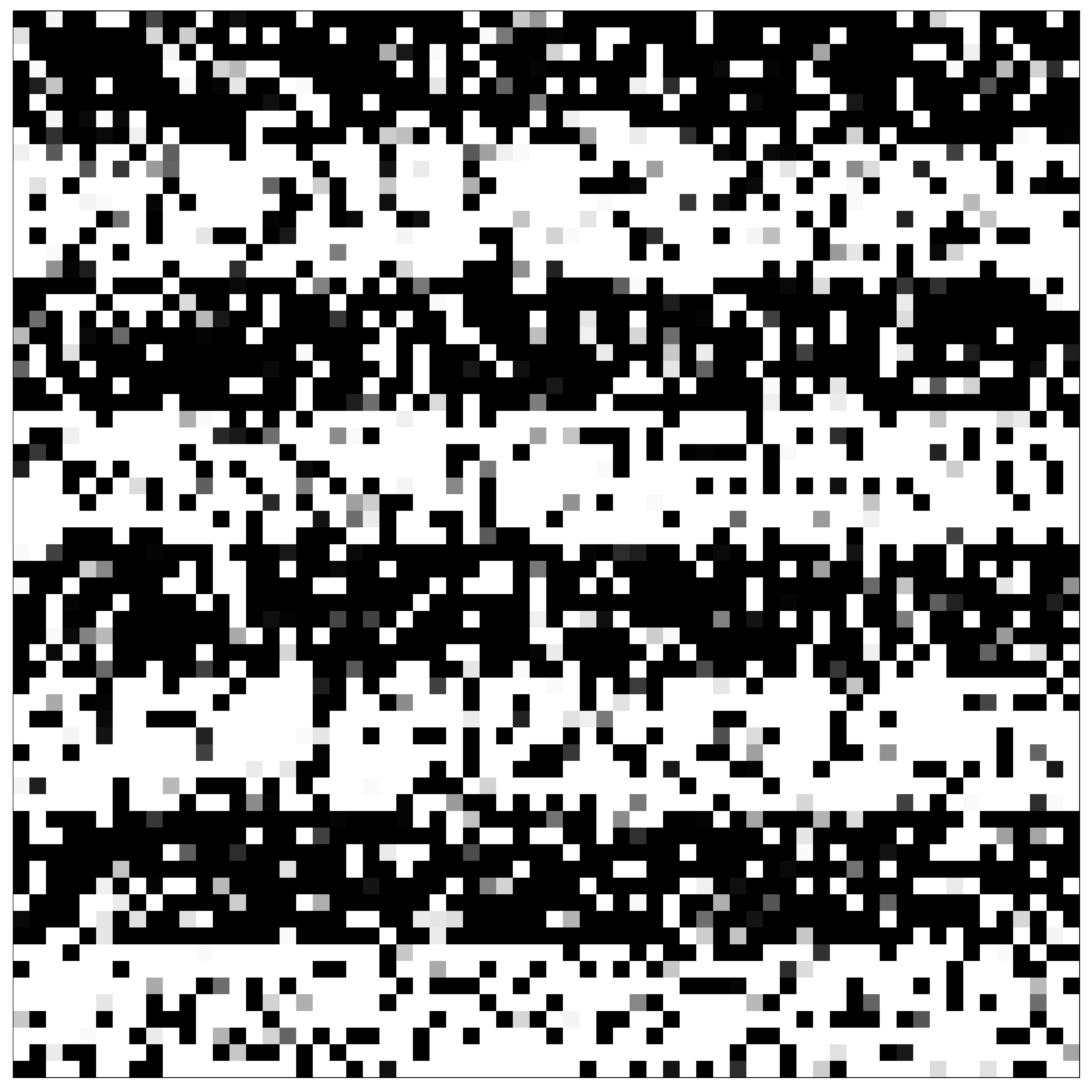}
  }
  \caption{\footnotesize A single SRAM cell and whole-SRAM power-on state.
  }
  \label{fig:eh_diagram_and_waveform}
  \hrulefill
\end{figure}
\label{sec:sram_aging}
The typical implementation of SRAM, which makes up main memory in embedded systems, caches in desktop-class processors, and configuration memory in Field-Programmable Gate Arrays, is two mutually reinforcing Complementary Metal-Oxide-Semiconductor inverters as shown in Figure~\ref{fig:sram_cell}.
These inverters form a bistable circuit where each state corresponds to a binary 1 or 0.
The power-on state of each cell is determined by a race condition between transistors M\textsubscript{2} and M\textsubscript{4}---as supply voltage increases, the transistor with the lower threshold voltage V\textsubscript{th} begins conducting earlier and drives the cell into one of the two stable states.
In contrast to Dynamic RAM, the mutually reinforcing inverters mean SRAM retains its state, without a refresh, as long as the cell is powered.

In a new device, V\textsubscript{th} is determined by manufacture-time processes that vary between each transistor causing a threshold voltage mismatch between the devices.
We identify three distinct sources of V\textsubscript{th} variation both within and across cells.
\begin{itemize}
    \item[] \textbf{Structural variation:} Design-time decisions such as wire length or choice of SRAM generator produce broad, family-level trends that are consistent and predictable across devices. We explore this source of variation in Section~\ref{sec:new_sram}.
    \item[] \textbf{Systematic variation:} Manufacturing variation stemming from wafer position or differences due to consistent operational offsets cause V\textsubscript{th} to vary. However, the tightly-coupled nature of SRAM ``filters out'' this V\textsubscript{th} variation in individual SRAM cells, as both inverters are affected equally. As a result, systematic variation has little impact on the V\textsubscript{th} mismatch within a cell, thus systematic variation has little influence on the SRAM power-on state statistics across devices.
    \item[] \textbf{Chaotic variation:} Local, transistor-scale physical processes such as random dopant fluctuation~\cite{mosfet-variation} contribute to V\textsubscript{th} mismatch by introducing random differences in the transistors within a cell. These random processes make the V\textsubscript{th} mismatch within an individual cell unpredictable.
\end{itemize}
When present, structural variation dominates V\textsubscript{th} in most cells and produces spatially correlated V\textsubscript{th} mismatches; however, the mismatch due to random variation in some cells is enough to counteract the structural variation.
Systematic variation affects all transistors in a cell equally and thus has little difference on a cell's V\textsubscript{th} mismatch.
We explore the interaction between structural and random biases, and their effect on SRAM power-on state, in Section~\ref{sec:new_sram}.

Threshold voltage differences between a cell's inverters dictate the power-on behavior of a cell;
the larger the difference, the more likely the cell is to power-on into the state associated with the lower V\textsubscript{th} inverter.
For most SRAM-based memories, most cells have relatively large V\textsubscript{th} mismatches and are thus ``strongly biased'' (\ie{} they always power-on to the same value).
For the few that are weakly biased, the V\textsubscript{th} mismatch between inverters is small enough that operational noise during the supply voltage ramp-up determines the cell's power-on state.
%Past work uses strongly biased cells as a source of device identification in the form of Physical Unclonable Functions (PUFs)~\cite{sram-puf-rng, microcontroller-sram-puf} and weakly biased cells as a source of random number generation~\cite{sram-puf-rng}.
We refer to the bias for a cell caused by manufacturing-time V\textsubscript{th} mismatch as hardware bias.

As devices age with use, transistor wear affects the V\textsubscript{th} mismatch;
with enough V\textsubscript{th} change, the effects are exposed in the cell's power-on state.
The dominant aging effect in SRAM is Bias Temperature Instability (BTI)~\cite{bti-in-sram, sram-bti-hci, nbti-digital-performance, nbti-digital-reliability}, which increases V\textsubscript{th} in the conducting (``on'') transistor.
% Checked, \cite{sram-bti-hci} says both BTI and HCI increase V\textsubscript{th}.
Two forms of BTI affect SRAM cells: Positive BTI degrades V\textsubscript{th} of the conducting n-type MOS (grounding), while Negative BTI (NBTI) degrades V\textsubscript{th} of the conducting p-type MOS (charging).
NBTI is of the most interest as it has a larger effect than PBTI and its net effect is to slow the $0$ to $1$ transition, which is the key factor in determining who wins the power-on race condition.

NBTI affects SRAM asymmetrically: when the SRAM cell is in a stable state, only one inverter conducts and is subjected to aging effects.
As V\textsubscript{th} rises for the transistor in the conducting inverter, that inverter is less likely to ``win'' the power-on race.
Because the bit stored in the cell determines the conducting inverter, aging is \textit{data-directed}: a cell aged holding a $1$ value is more likely to power-on to a $0$ on the next power cycle, and vice versa.
Past work demonstrates SRAM's susceptibility to adversarial, \textit{intentional} asymmetric aging by using it to clone or destroy SRAM PUFs~\cite{cloning-pufs, attacking-puf}, as well as recover secret keys~\cite{challenging-sram-security}.
\textbf{This paper leverages the software-directed aging that occurs naturally over a device's lifetime, referred to further as software bias, to distinguish between new and recycled devices.}

\section{Motivation}
\label{sec:motivation}

We exploit software-directed SRAM aging to detect recycled devices \textit{without enrollment, modification, or destruction} by measuring the power-on state statistics of a device's SRAM.
Three observations drive our approach: (1) SRAM's power-on state statistics are highly-regular across devices, (2) SRAM power-on states change deterministically due to software-directed aging, and (3) software has innate data asymmetries that gradually imprint on SRAM's analog domain, revealed digitally through biases in SRAM's power-on states.
We experimentally validate each of these observations before leveraging them to design our recycled counterfeit device detector.

\subsection{New SRAM is Regular and Predictable}
\label{sec:new_sram}

% Version with 2 gecko boards.
\begin{table*}[t]
\centering
\footnotesize
\begin{tabular}{l | c c c | c c c | c c c}
\textbf{Device}           & \multicolumn{3}{c|}{MSP430G2553~\cite{g2553datasheet}}    & \multicolumn{3}{c|}{MSP432P401R~\cite{msp432datasheet}}   & \multicolumn{3}{c}{EFM32WG990F256~\cite{efm32datasheet}} \\
\hline
\textbf{Manufacturer}   & \multicolumn{3}{c|}{Texas Instruments}                    & \multicolumn{3}{c|}{Texas Instruments}                    & \multicolumn{3}{c}{Silicon Labs}     \\
\textbf{Core}           & \multicolumn{3}{c|}{MSP430}                               & \multicolumn{3}{c|}{ARM Cortex-M4}                        & \multicolumn{3}{c}{ARM Cortex-M4}    \\
\textbf{SRAM Size (B)}  & \multicolumn{3}{c|}{512}                                  & \multicolumn{3}{c|}{64K}                                  & \multicolumn{3}{c}{32K}              \\
\textbf{Aging Acceleration} & \multicolumn{3}{c|}{Voltage, Temperature}             & \multicolumn{3}{c|}{Temperature}                          & \multicolumn{3}{c}{Temperature}      \\
\textbf{Dev. Board Cost} & \multicolumn{3}{c|}{\$9.99}                           & \multicolumn{3}{c|}{\$19.99}                                & \multicolumn{3}{c}{\$99.99}            \\
\textbf{Sample size}    & \multicolumn{3}{c|}{18}                                   & \multicolumn{3}{c|}{6}                                    & \multicolumn{3}{c}{2}          \\
\hline
\textbf{Statistic}  & Mean & Std. Dev. & Rel. Std. Dev.           & Mean & Std. Dev & Rel. Std. Dev.     & Mean & Std. Dev & Rel. Std. Dev. \\
\hline
Mean bias               & 51.9\%  & 1.5\% & 2.89\%       & 49.5\%  & 0.05\% & 0.11\%        & 46.8\% & 0.69\% & 1.49\% \\
Portion strong          & 88.4\% & 1.1\% & 1.24\%       & 78.5\% & 0.76\% & 0.97\%        & 77.5\% & 4.72\% & 6.09\% \\
Portion weak            & 11.6\% & 1.1\% & 9.48\%       & 21.5\% & 0.76\% & 3.54\%        & 22.5\% & 4.72\% & 21.0\% \\
Weak bias mean          & 49.9\%  & 1.6\% & 3.14\%       & 50.0\% & 0.06\%  & 0.13\%        & 53.2\% & 3.3\% & 6.12\% \\
%Weak bias std. dev.     & 0.355  & 0.004 & 1.24\%       & 0.371 & 6.6E-4  & 0.18\%        & 0.390 & 0.017 & 4.41\% \\
Portion strong 1        & 46.1\% & 1.2\% & 2.60\%       & 38.8\% & 0.35\% & 0.89\%        & 34.7\% & 3.93\% & 11.3\% \\
Portion strong 0        & 42.3\% & 1.9\% & 4.49\%       & 39.7\% & 0.42\% & 1.05\%        & 42.9\% & 0.79\% & 1.84\% \\
Spatial autocorrelation               & 0.243  & 0.057 & 24.3\%       & 0.022  & 0.010  & 48.3\%        & 0.0015 & 0.0066 & 435\% \\
$p$-value*    & 0      & 0     & -            & 0      & 0      & -             & 0.001 & 0 & 0 \\
\end{tabular}
\caption{\footnotesize SRAM power-on statistics for devices from three different microcontroller families. * $p$-value for spatial autocorrelation; Python 3.6.9 rounds the $p$-value down to 0 for the MSP430 and MSP432.}
\label{table:new_stats}
\hrulefill
\end{table*}

\begin{table}[!t]
\centering
\footnotesize
\begin{tabular}{l|c c c c}
% \hline
\textbf{Change}     & Mean      & Std. Dev.     & Rel. Std. Dev & $p$-value\\
\hline
Mean Bias           & 0.088     & 0.0004        &  0.5\%    &  1.8E-19\\
Portion strong      & -1.08\%   & 0.26\%        & 23.7\%    &  0.00023\\
Portion weak        & 1.08\%    & 0.26\%        & 23.7\%    &  0.00044\\
Weak bias mean      & -0.026    & 0.020         & 24.1\%    &  0.01625\\
Weak bias std. dev. & 0.0047    & 0.0045        & 95.7\%    &  0.00554\\
Portion strong 1    & 8.56\%    & 0.20\%        & 2.29\%    &  1.8E-22\\
Portion strong 0    & -9.64\%   & 0.26\%        & 2.70\%    &  3.0E-16\\
\end{tabular}
\caption{\footnotesize Change in summary statistics after fully biased aging towards 1 on the MSP430G2553.}
% Source: change_distribution.py
\label{table:change_stats}
\hrulefill
\end{table}

% Including towards-0 and towards-1
% \begin{table}[!t]
% \centering
% \footnotesize
% \begin{tabular}{l|c c c | c c c}
% % \hline
%  & \multicolumn{3}{c|}{Towards 0} & \multicolumn{3}{c|}{Towards 1} \\
% \textbf{Change}     & Mean      & Std. Dev.     & Rel. Std. Dev.    & Mean      & Std. Dev.     & Rel. Std. Dev.\\
% \hline
% Mean Bias           & x.xxx     & x.xxx         & x.xx\%            & x.xxx     & x.xxx         & x.xx\%\\
% Portion strong      & xx.x\%    & x.x\%         & x.xx\%            & x.xxx     & x.xxx         & x.xx\%\\
% Portion weak        & xx.x\%    & x.x\%         & x.xx\%            & x.xxx     & x.xxx         & x.xx\%\\
% Portion strong 1    & xx.x\%    & x.x\%         & x.xx\%            & x.xxx     & x.xxx         & x.xx\%\\
% Portion strong 0    & xx.x\%    & x.x\%         & x.xx\%            & x.xxx     & x.xxx         & x.xx\%\\
% \end{tabular}
% \caption{\footnotesize Change in summary statistics after fully biased aging on the MSP430G2553.}
% % Source: flipping_stats.py
% \label{table:change_stats}
% \hrulefill
% \end{table}

\begin{figure}[ht]
  \centering
  \includegraphics[width=0.8\columnwidth]{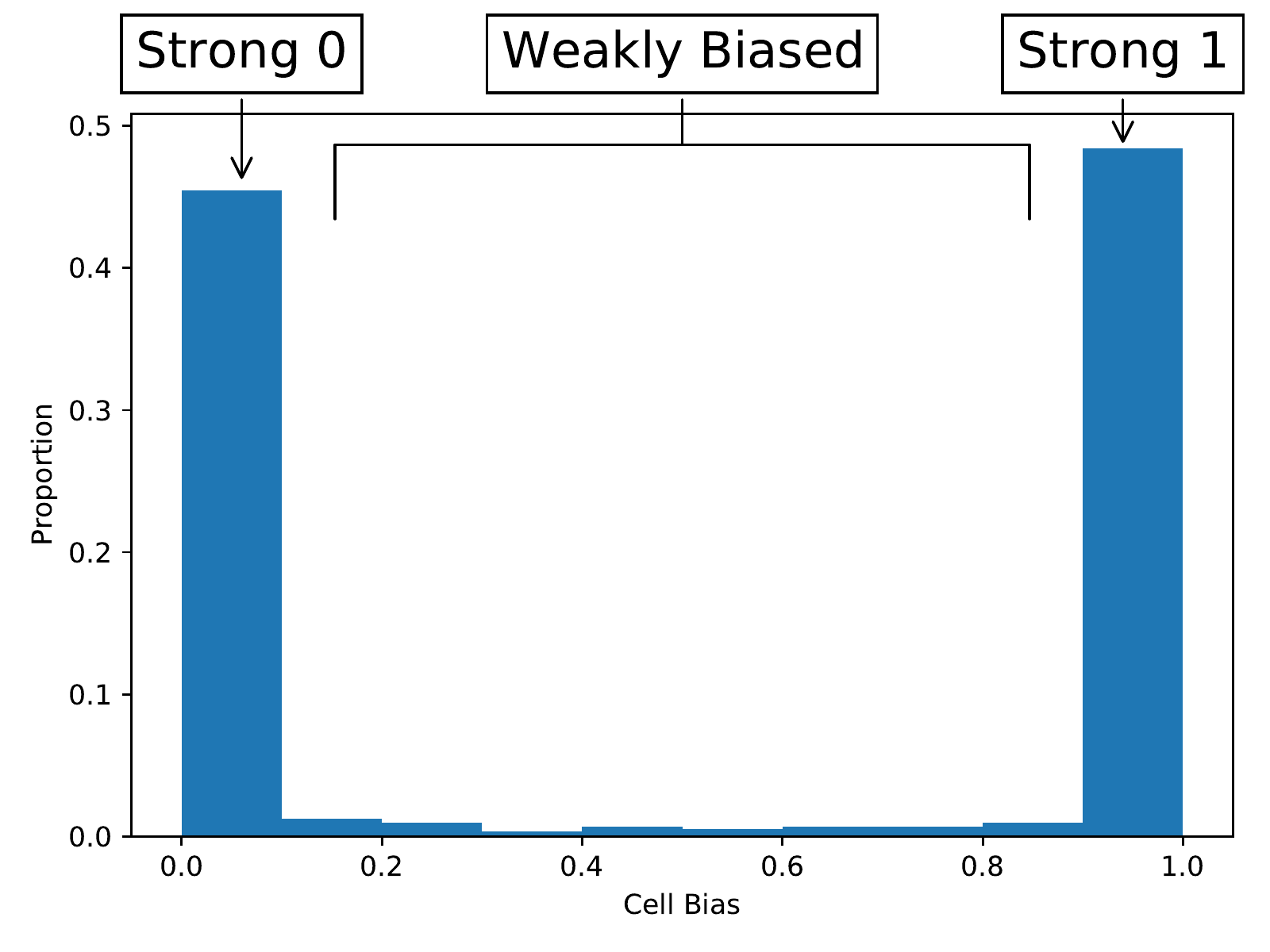}
  \caption{\footnotesize Distribution of cell biases in the baseline MSP430 set.}
  \label{fig:bias_histogram}
  \hrulefill
\end{figure}

The first observation behind our approach is that, unlike individual cells, which are influenced by chaotic variation, whole-SRAM power-on state statistics are very similar across different devices of the same type.
This makes sense when considering that devices contain many thousand SRAM cells that serve to average-out the effects of chaotic variation when taken together.
To quantitatively support this observation, we measure the inter-device variation of whole-SRAM power-on state statistics for three microcontrollers (details in Table~\ref{table:new_stats}): MSP430, MSP432, and EFM32.
We choose these devices for variety in both manufacturer (Texas Instruments manufactures the MSP430 and MSP432, while Silicon Laboratories manufactures the EFM32) and processor (the MSP432 and EFM32 are ARM Cortex-based microcontrollers and the MSP430 uses its own low-cost, low-power core), and because they are represent the most popular legacy and current microcontrollers.

We determine the bias of the cells in each device by recording SRAM's power-on state 51 times at 20\degree{}C.
We select 20\degree{}C to limit the impact of thermal noise and aging on our power-on state measurements.
We use a large sample size to reduce the impact of randomness on our measurements and to get a fine-grain notion of each cell's power-on probability.
Using an odd number of samples enables us to determine a majority value for each cell.

We find that while the biases of individual cells are unpredictable across devices in a design due to random chaotic variation, the high-level representative statistics of new SRAM are highly predictable.
Table~\ref{table:new_stats} shows several summary statistics for the SRAM power-on states on each microcontroller.
We separate SRAM cells into one of three possible categories:
\begin{itemize}
    \item[] \textbf{Strongly $1$-biased:} Cells that power-on to $1$ every trial.
    \item[] \textbf{Strongly $0$-biased:} Cells that power-on to $0$ every trial.
    \item[] \textbf{Weakly biased:} Cells that power-on into both states over the course of the trials, \eg{} a cell powering-on into $0$ 40 times and $1$ 11 times.
\end{itemize}
Figure~\ref{fig:bias_histogram} shows the proportion of strong and weak cells in the MSP430s.
The power-on probability follows a symmetric U-shaped distribution where the strongly-biased cells dominate.
For each new device within a each the same device type, the portion of strongly $0$- and $1$-biased cells as well as the mean bias of SRAM is consistent.
Different designs have different degrees of similarity between devices: the MSP432 devices are highly regular, while the MSP430 and EFM32 vary more between devices.
% While these statistics allow us to form an accurate expectation for new devices, their value as an indicator of device age depends on both how they change with age and the variation of that change between similarly-aged devices.

We evaluate the effect of IC-level design choices on cell biases in the form of spatial correlation, shown in Table~\ref{table:new_stats}.
We calculate spatial autocorrelation using Moran's I~\cite{moran, pysal} in order to quantify patterns in cell biases like the alternating-stripe pattern in Figure~\ref{fig:preage_heatmap}.
Moran's I varies between -1 and 1, where values further from 0 imply a non-random spatial pattern and the direction of the value indicates a negative correlation or a positive correlation.
A zero value indicates no spatial autocorrelation: values are randomly spread throughout the space.
We quantify the strength of cell bias patterns caused by design-time choices using Moran's I and the $p$-value with a random distribution null hypothesis for each device family in Table~\ref{table:new_stats}.
% The MSP430 and MSP432 both contain a non-random pattern in cell biases; the pattern is more evident in the MSP430 as shown by the higher value for Moran's I.
% Not all designs exhibit significant spatial autocorrelation: none of the EFM32 devices under test display a predictable pattern in cell biases.
Each device family tested contains a non-random pattern in cell biases; however, the strength of the pattern varies---spatial correlation is more evident in the MSP430 than in the MSP432 or EFM32, as shown by the higher value for Moran's I.
The more significant the spatial autocorrelation, the more information it provides during classification, as discussed in Section~\ref{sec:structural-awareness}.

\subsection{SRAM Aging is Regular and Predictable}
\label{sec:old_sram}

The low deviation of the values in Table~\ref{table:new_stats} enable an accurate statistical model for new devices, but their value as indicators of device age depends on both how they change with age and the variation of that change between similarly-aged devices.
To quantify how the values in Table~\ref{table:new_stats} change with age, we age the embedded SRAM within a set of six MSP430G2553s for an effective five years using the experimental setup described in Section~\ref{sec:setup}.
We choose the MSP430G2553s primarily because they expose the unregulated SRAM supply voltage, which enables us to observe the long-term effects of device aging in a short time.\footnote{We conduct aging experiments analogous the one presented in this section using the MSP432P401R for two months (3.6 effective months given the 1.8x acceleration factor provided by increasing temperature from 293 K to the max operating temperature of 358 K). We observe similar cell bias shift trends in both, but focus on the MSP430 to explore aging's long-term effects.}
Five years captures the majority of SRAM's change due to age: our evaluation in Sections \ref{sec:bounding-aging} and \ref{sec:operating-time} indicates that change due to age follows a roughly logarithmic trend and that most measurable change occurs in the first 3-4 months of use, consistent with trends observed in past work~\cite{ro-aging, fpga-aging}.

In order to determine how the data in a cell affects the magnitude and direction of its aging, we age half of the devices by writing all $1$s to the SRAM and half of the devices writing all $0$s.
Table~\ref{table:change_stats} details the change in summary statistics we observe in the devices aged with all $0$s (\ie{} all cells aged towards a power-on state of $1$).
Note that the change in the set of all-$1$ devices is similar, but in the opposite direction---the bit stored in the cell determines the direction, but not the magnitude, of the bias change.
For each statistic, Table~\ref{table:change_stats} also includes $p$-values as calculated by Welch's t-test~\cite{welch} after the observed change is applied to samples of the new distribution.
The null hypothesis is that the resulting statistical values are samples of the new-device statistical distribution, \ie{} the $p$-value represents how likely it is to find a new device with such a statistic, given the new device distribution for that statistic.
We use the $p$-value as a heuristic for the usefulness of the statistic in classification (smaller is better, because it indicates increased new and aged device separation).
These results reveal two key insights about SRAM aging:
\begin{enumerate}
    \item[] \textbf{SRAM aging is \textit{predictable}}: similar devices respond in similar ways to age-based wear, which enables accurate modeling of SRAM's statistical changes as devices age.
    \item[] \textbf{SRAM aging is \textit{distinguishing}}: the change induced by aging is much larger than the variation expected in new devices, which creates a clear distinction between new and used devices.
\end{enumerate}

Because SRAM ages equally in either direction (towards $0$ or $1$), the primary factor in the rate of software burn-in is the bias of the bit software writes to the cell.
Unbiased bits that spend roughly equal time containing as $0$ or $1$ do not change SRAM power-on state statistics, as both inverters age equally and the V\textsubscript{th} mismatch within the cell stays constant.
Thus, SRAM power-on states only reveal device age when software uses the SRAM asymmetrically.

\subsection{Software is Naturally Biased}
\label{sec:software_asymmetry}
\begin{figure}
\centering
\subfloat[PID bias]
{
    \includegraphics[width=.47\columnwidth]{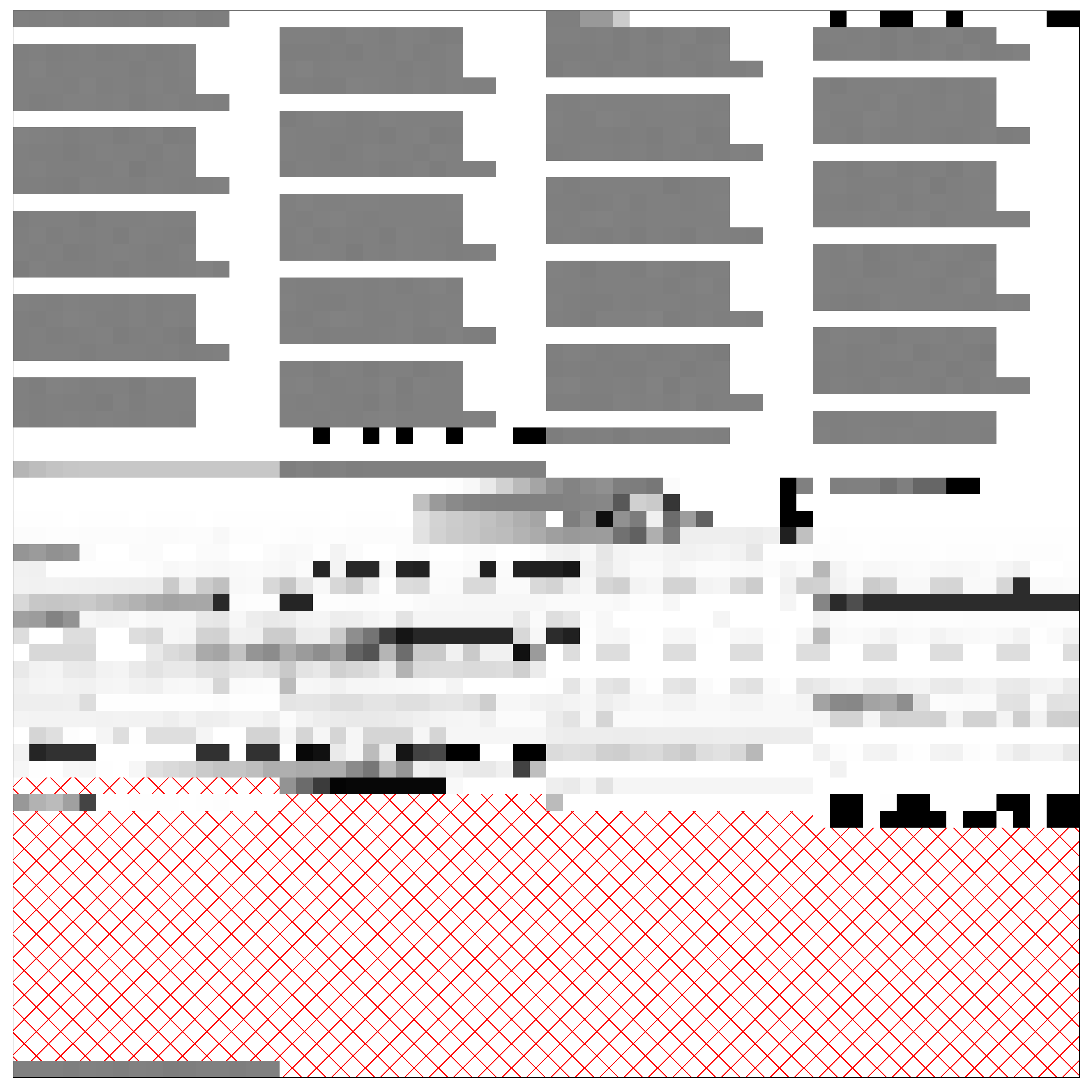}
    \label{fig:pid_bias}
}
\subfloat[FFT bias]
{
    \includegraphics[width=.47\columnwidth]{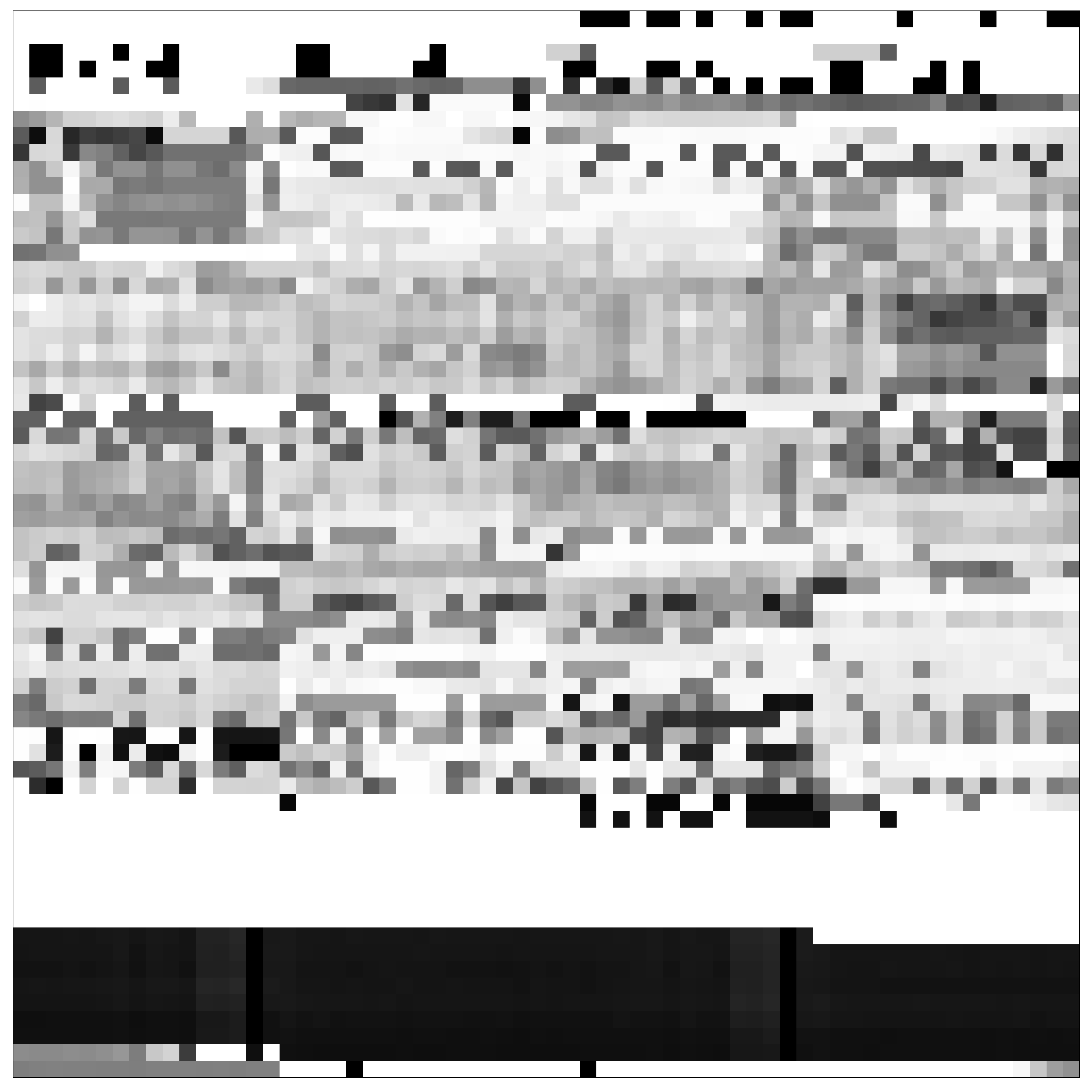}
    \label{fig:fft_bias}
}
\caption{\footnotesize Bit biases for the PID and FFT benchmarks. A darker square indicates that bit spent more time as a 1; hatching indicates that memory location is never written by the benchmark.}
\label{fig:pid_and_fft_bias}
\hrulefill
\end{figure}
\begin{table}[t]
\footnotesize
\centering
\begin{tabular}{l c c c}
\textbf{Benchmark}            & \textbf{SRAM Use (B)} & \textbf{Avg. SW Bias} & \textbf{Mean Strength}\\
\hline
\hline
FFT                 & 512   & 0.3074    &    0.6897\\
FSM Controller      & 468   & 0.0744    &    0.8687\\
FIR Filter          & 448   & 0.3356    &    0.3957\\
Quicksort           & 512   & 0.1995    &    0.7738\\
PID Controller      & 384   & 0.2190    &    0.6228\\
% String Generator    & 447   & 0.3493    &    0.4698\\
\textbf{Average}    &       & 0.2272    &    0.6701\\
\end{tabular}
\caption{\footnotesize Benchmarks that we use to test \sys{}.}
% Source: sim_bias_analysis.py
\label{table:benchmarks_combined}
\hrulefill
\end{table}

% \begin{table}[!t]
% \footnotesize
% \centering
% \begin{tabular}{l c c c c}
% \textbf{Benchmark}            & \textbf{Code Size (B)} & \textbf{SRAM Use (B)} & \textbf{Avg. SW Bias} & \textbf{Mean Bit Strength}\\
% \hline
% \hline
% FFT                 & 16212     & 512   & 0.3074    &    0.6897\\
% FSM Controller      & 1542      & 468   & 0.0744    &    0.8687\\
% FIR Filter          & 4240      & 448   & 0.3356    &    0.3957\\
% Quicksort           & 3678      & 512   & 0.1995    &    0.7738\\
% PID Controller      & 9714      & 384   & 0.2190    &    0.6228\\
% String Generator    & 1024      & 447   & 0.3493    &    0.4698\\
% \end{tabular}
% \caption{\footnotesize Benchmarks that we use to test \sys{}.}
% % Source: sim_bias_analysis.py
% \label{table:benchmarks}
% \hrulefill
% \end{table}
Our second driving observation is that embedded software intrinsically contains asymmetric data patterns.
Specifically, most memory cells spend more time holding a $0$ rather than a $1$.
% Past work observes that the ratio of 1s to 0s in regular system files is well below 50\%~\cite{file-hamming}; we extend this observation and show that the same is true for main memory used by software.
We validate this claim using five embedded benchmarks representing common software tasks, described in Table~\ref{table:benchmarks_combined}.
We generate pseudorandom input data for each benchmark using a Linear Feedback Shift Register (LFSR)~\cite{lfsr} to ensure biased input data does not affect the results.
% Each benchmark generates its own pseudorandom input data using a Linear Feedback Shift Register (LFSR)~\cite{lfsr} to ensure repeatability and predictability in testing.

We measure software asymmetry by running each benchmark in a modified version of the MSP430 simulator included in the open source software MSPDebug~\cite{mspdebug} and track the portion of total time each SRAM cell contains a certain value over the course of a large number (10,000) of executions.
Figure~\ref{fig:pid_and_fft_bias} illustrates the results for the \texttt{PID} and \texttt{FFT} benchmarks as a heatmap of bit\footnote{Throughout this paper we make a distinction between \textit{bit bias}, the time-averaged value of a software bit during execution, and \textit{cell bias}, the proportion power-ons a SRAM cell takes on a $1$ value. For example, a bit with a bias of 0.8 holds $1$ for 80\% of execution time, while a cell with a bias of 0.8 powers on to $1$ in 80\% of power-on trials.} biases.
Darker values indicate that the bit spent more of the total benchmark time containing a $1$.
Across 10,000 executions of each of the five benchmarks, we find that the mean bit bias is 0.227---that is, any given bit in program memory spends on average 22.7\% of the total execution time containing a $1$.
% Source: benchmark_flips.py
Taken with the results of Section~\ref{sec:new_sram} that SRAM ages towards powering-on as the inverse of its held value, we observe that \textbf{the number of cells biased towards powering into the $1$ state increases over time}.

While mean bit bias describes the final state of the SRAM given sufficient aging time, it does not accurately represent the rate at which software burn-in occurs.
For example, a program in which 75\% of bits always contain $1$ and 25\% always contain $0$ burns into SRAM faster than a program in which all bits contain a $1$ 75\% of the time and a $0$ 25\% of the time, despite both programs having the same mean bias of 0.75.
In order to better distinguish between programs that age hardware at different rates, we quantify the bias strength of a bit as $Strength = 2 * |B - 0.5|$, where $B$ is the bias of the bit.
We use bias strength as a predictor of how likely a software bit is to change a cell's power-on statistics (thereby providing age information), and extend it to software in the form of mean bit strength.
We report the average bit strength for each benchmark in Table~\ref{table:benchmarks_combined}.
The results indicate that software includes sufficient strongly-biased bits to change cell power-on states.
Taken together, the mean bias and bit strength quantify the magnitude, direction, and rate of software burn-in.

\section{Design}
\begin{figure*}[t]
  \includegraphics[width=2\columnwidth]{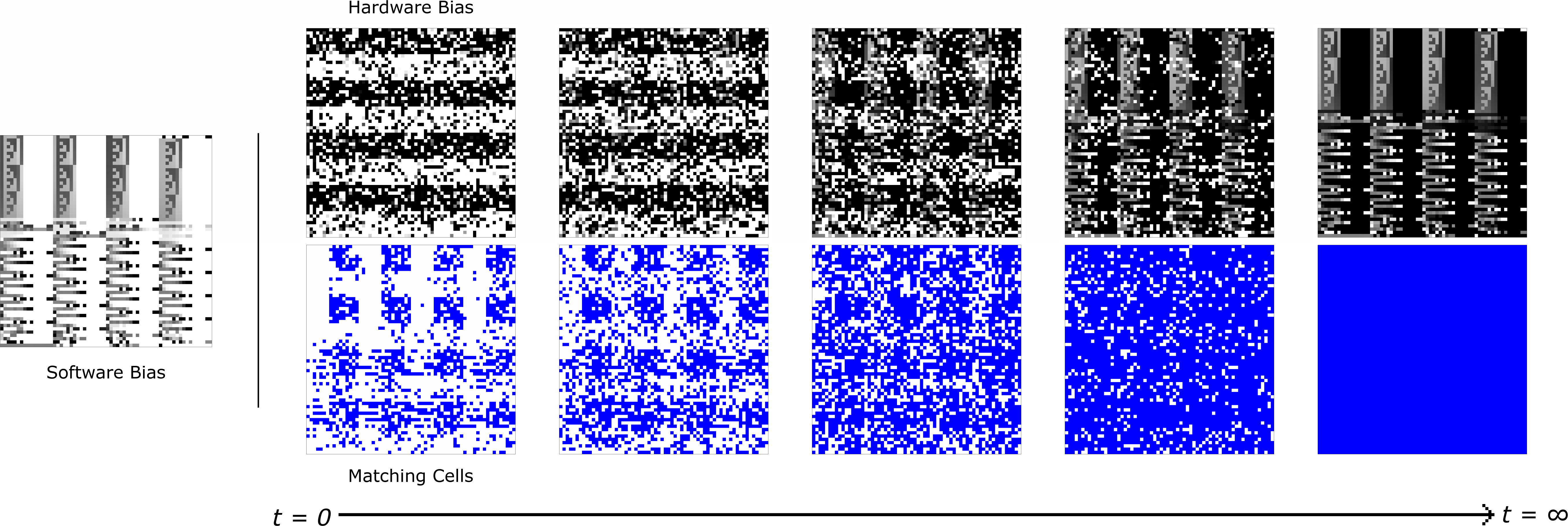}
  \caption{\footnotesize Gradual software burn-in to hardware.}
  \label{fig:burn-in}
  \hrulefill
\end{figure*}

To leverage software's distinctive, asymmetric aging effect on SRAM as illustrated in Figure~\ref{fig:burn-in}, we design \sys{}: a tool that detects recycled computing devices.
\sys{} quantifies software-induced bias in SRAM's power-on state through whole-SRAM statistics.
\sys{} then compares the resulting bias to the expected statistical bias of a new device;
the degree of similarity determines whether the device is classified as new or recycled.

We build \sys{} around the abstraction of a noisy information channel.
We identify two noise sources: (1) SRAM cell aging visibility (i.e., whether aging impacts an SRAM cell sufficiently to change its power-on probability) and (2) software bit bias (i.e., whether a bit holds a single value the majority of the time).
These noise sources serve to limit the amount of information available through power-on state measurements and are not under our control.
We identify three sources of information, each assuming a different amount of software- and device-specific knowledge:
(1) universal trends: features that are consistent across software and devices (e.g., new device mean bias is ~50\%);
(2) structure awareness: whole-SRAM features seen across devices (of the same type) (e.g., the MSP430 has an alternating 512-bit stripe pattern);
and (3) software awareness: software-specific features (e.g., the regions of strongly 1- and 0-biased bits near the bottom of the FFT benchmark memory).
By modulating access to these information sources, we create use-case-specific expectations for new and recycled devices.
We then classify devices as new or recycled using these expectations.

\subsection{Zero-knowledge Classification}
\label{sec:unaware-design}

The most general use-case for \sys{} is when no device- or software-specific information is available.
In such cases, the only sources of information to use in classifying a device are general trends seen across devices and software.
As Section~\ref{sec:motivation} shows, there are two general trends that we identify: (1) embedded software is inherently biased towards 0 and (2) SRAM embedded in new computing devices tends to have an equal number of $0$'s and $1$'s in its power-on state.
The first trend forms our aged device expectation: we assume that all software bits stored in SRAM cells are $0$; this means we expect SRAM power-on states to age towards an increased number of $1$s.
The second trend forms our new device expectation: we assume an initial 50/50 balance of $1$s and $0$s in the whole-SRAM power-on state.
Thus, aging serves to push a device's power-on state from 50\% $1$s towards 100\% $1$s.

Based on the $p$-value heuristic shown in Table~\ref{table:change_stats}, the best potential metrics for detecting recycled devices in this scenario are (1) the mean cell bias, (2) the count of strongly 0-biased cells, and (3) the count of strongly 1-biased cells.\footnote{The statistics for weakly-biased cells remains evenly distributed around 50\% before and after aging. This means that they do not add value as a summary statistic for classification purposes. We explore how weakly-biased cells reveal age through their location in Section~\ref{sec:structural-awareness}.}
Because strongly-biased cells dictate the mean bias,\footnote{Strongly-biased cells make up the majority (81.5\% on average between families and 88.4\% in the MSP430) of all cells and the mean bias of weakly-biased cells is unchanged due to aging.} the zero-knowledge classifier uses only the count of strongly-biased $0$ and $1$ cells. 
Putting it all together, we expect the average recycled device has significantly more strong-$1$ cells and significantly fewer strong-$0$ cells than a new device.
For classification, \sys{} counts the strong-$1$ and strong-$0$ cells and generates a score $S = count_{1} - count_{0}$; if the score is above a threshold value $T$, the device is marked as recycled.

\subsection{Structure-aware Classification}
\label{sec:structural-awareness}

\begin{figure}[t]
  \includegraphics[width=\columnwidth]{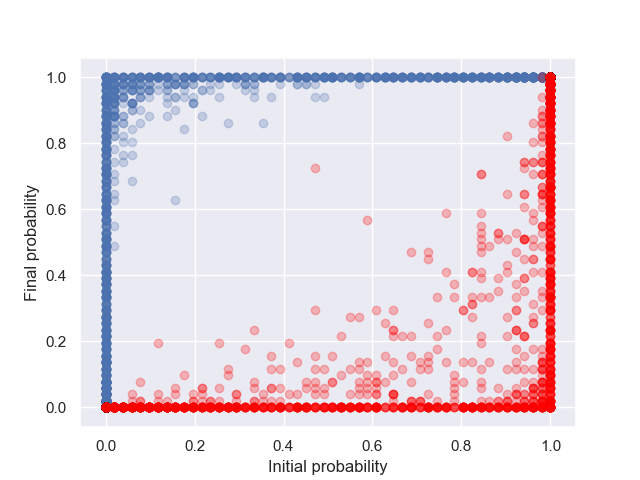}
  \caption{\footnotesize Relationship between cell initial bias and bias after aging.}
  \label{fig:aging-before-after}
  \hrulefill
\end{figure}

While the zero-knowledge classifier leverages broad truths about hardware and software biases, we observe that devices of the same type are subject to structural variation that results in biases in SRAM's power-on state that span devices.
We leverage those cross-device biases to enhance \sys{} through structural awareness.
For example, any new SRAM taken as a whole is characterized by a roughly even split of $0$ and $1$ cells, but asymmetries in the device's design can result in spatial locality (quantified by Moran's I in Section~\ref{sec:new_sram}) between cell power-on states.
Take the MSP430G2553 shown in Figure~\ref{fig:preage_heatmap}; the structural bias is obvious in the alternating 512-cell bands of cells biased unexpectedly towards $0$ or $1$, compared to the 51.9\% seen at the whole-SRAM level.
Quantitatively, we see that, for a given band, ~68\% of cells are strongly-biased towards the majority value, ~12\% of cells are weakly-biased, and ~20\% of cells are strongly-biased towards the minority value.

Next, consider the results of extreme software aging shown in Figure~\ref{fig:aging-before-after}.
These results indicate that weakly-biased cells in new devices are highly susceptible to aging and are very likely to become strongly-biased cells in recycled devices.
Quantitatively, 88\% of weakly-biased cells in the new MSP430s that we tested became strongly-biased cells in recycled devices, for both software extremes.
Notice that the number of weakly-biased cells stays the same (at the whole-SRAM level); this is because the weakly-biased cells that change to strongly-biased cells are replaced by initially strongly-biased cells that become weakly-biased cells.

At the band level we see a different trend: when you age away from the majority value of a band, there are many more cells that become weakly-biased cells, compared to aging towards the majority value.
Mathematically, this makes sense as 68\% of cells in a band are strongly-biased towards the majority, while only 20\% of the cell in a band are strongly-biased towards the minority.
Since 18\% of initially strongly-biased cells become weakly biased when aged against their bias, aging away from the majority results in 12.2\% of a band's cells becoming weakly-biased, while aging towards the majority results in 3.6\% of a band's cells becoming weakly-biased.
Thus, we make the high-level conclusion that \textbf{aging away from a biased region's majority value increases the number of weakly biased cells, while aging towards the majority value decreases it}.

\sys{} exploits this distinction to gain some information from weakly-biased cells.
Specifically, \sys{} considers a weakly-biased cell a match to the expected value (given the corresponding software bit value) only if it is in a band whose majority value is the opposite of the expected value; otherwise, the weakly-biased cell is marked as a disagreement.
Table~\ref{table:scoring} covers all cases of expected value, band majority value, cell bias, cell value, and whether \sys{} considers the combination to be a match or a disagreement.
The total score $S$ for a devices is the number of matches minus the number of disagreements across all cells in the device's SRAM: $S = count_{+} - count_{-}$.
Notice that this algorithm naturally accounts for changes in structural bias.
As structural bias approaches 50\% (i.e., none), the aging towards and away from the majority values produces more similar numbers of weakly-biased cells and the $+$s and $-$s due to weakly-biased cells tends to cancel-out.
This performs similar to the zero-knowledge classifier, except there we ignore weakly-biased cells, because they act as noise.
As structural bias increases towards 100\%, there are no weakly-biased cells created by aging towards the majority, hence no $-$s due to weakly-biased cells;
only $+$s are possible from weakly-biased cells due to aging away from the majority.
Thus, \textbf{our structural-aware classifier supplants the zero-knowledge classifier---regardless of the magnitude of structural bias}.

% Proportion of total bits in a single band is 1.3\% increased for away and 2.0 decreased for towards.
% Start with 12\% of 512 = 61 weak bits
% Go up to 13.3\% of 512 = 68 weak bits
% Go down to 10\% of 512 = 51 weak bits
% 68 / 61 = 1.115 -> 12\% increase
% 51 / 61 = 0.836 -> 16\% decrease

\subsection{Software-aware Classification}
\label{sec:aware-design}

\begin{table}[!t]
\centering
\footnotesize
\begin{tabular}{l|c|c|c|c|c|c|}
% \hline
\cline{2-7}
 & \multicolumn{3}{c|}{\textbf{0 Majority Band}} & \multicolumn{3}{c|}{\textbf{1 Majority Band}} \\
% \backslashbox{\textbf{Expected}}{\textbf{Actual}} & \makebox{\textbf{0}} & \makebox{\textbf{Weak}} & \makebox{\textbf{1}} & \makebox{\textbf{0}} & \makebox{\textbf{Weak}} & \makebox{\textbf{1}}\\
\hline
\multicolumn{1}{|c|}{\backslashbox{\textbf{Expected}}{\textbf{Actual}}} & \textbf{0} & \textbf{Weak} & \textbf{1} & \textbf{0} & \textbf{Weak} & \textbf{1}\\
\hline
\multicolumn{1}{|l|}{\textbf{0}}          & + & - & - & + & + & -\\
\hline
\multicolumn{1}{|l|}{\textbf{1}}          & - & + & + & - & - & +\\
\hline
\end{tabular}
\caption{\footnotesize Per-cell scoring chart, based on expected value (software-dependent) vs. actual power-on value and strength.}
% Source: flipping_stats.py
\label{table:scoring}
\hrulefill
\end{table}

While the zero-knowledge classifier leverages the broad observation that embedded software tends to be $0$ heavy, software analysis provides an opportunity to exploit specific insights about a given piece of software (or set of suspect software).
\sys{} uses software information in the form of individual bit bias as discussed in Section~\ref{sec:software_asymmetry} to better inform the expected aged SRAM power-on distribution.
Software analysis improves \sys{} by replacing the core assumption that software is generally biased towards 0 with the \textit{ground truth} for the device running the known software: which software bits are strongly biased towards 0 or towards 1, and which bits are weakly biased or never written.
Based on the individual bit bias for each benchmark tested, we generate an expected power-on state distribution for a device aged for infinite time with that software (\ie{} the point when software bias completely dominates hardware bias).
The expected distribution for a given piece of software corresponds to the inverse of the bias distribution shown in Figure~\ref{fig:pid_and_fft_bias}.
For example, a software bit with a bias of $0$---the cell always contains a $0$---causes that cell to have an expected $1$-dominate power-on bias.

For the software-aware classifier, we separate software bits into three bins, based on a bias strength threshold (as calculated in Section~\ref{sec:software_asymmetry}): usable $1$-biased, usable $0$-biased, and unusable.
As discussed in Section~\ref{sec:software_asymmetry}, only bits sufficiently biased in one direction have the potential to change the power-on probability of their associated SRAM cell.
Choosing which bits/cells to examine is a balancing act: including weaker bits reduces the chance that the associated cell ages enough to be detectable, while setting the threshold to be too strict reduces the total information available to make the classification decision.
We explore the effect of varying the software bit threshold on classification ability in Section~\ref{sec:aware-eval}.
\sys{} also ignores cells the software does not write, as these cells age in a software-independent manner.
\sys{} calculates a score for the device based on the remaining cells using the rules in Table~\ref{table:scoring}, using the same algorithm used in structural-aware classification.\footnote{Structural and software awareness are orthogonal information sources.}

\subsection{Classification}

The classification decision for a Device Under Test (DUT) is based on the difference between its score at the time of testing and the expected score for new devices.
Because differences in cell bias result from random physical process variations, we model the possible scores for new devices as a normal distribution with a mean and variance based on the statistics in Table~\ref{table:new_stats}.
We consider the DUT recycled if its score is beyond a decision threshold $T$ number of standard deviations greater than the expected score for new devices.
The best $T$ is software-dependent, because different benchmarks have different levels of score variation for new devices and different levels of score change in aged devices.
$T$ is also application-dependent based on the relative costs of false positives and negatives; we explore this trade space for each configuration of \sys{} throughout Section~\ref{sec:evaluation}.

% Outstanding questions
%   Trends hold for other/newer devices
%   Adversarial countermeasures:
%     Impact of recovery
%     Impact of counter-aging
\section{Evaluation}
\label{sec:evaluation}
\label{sec:setup}

%Captions may need to be changed to a smaller font?
\begin{figure}[t]
\centering
  \includegraphics[width=0.7\columnwidth]{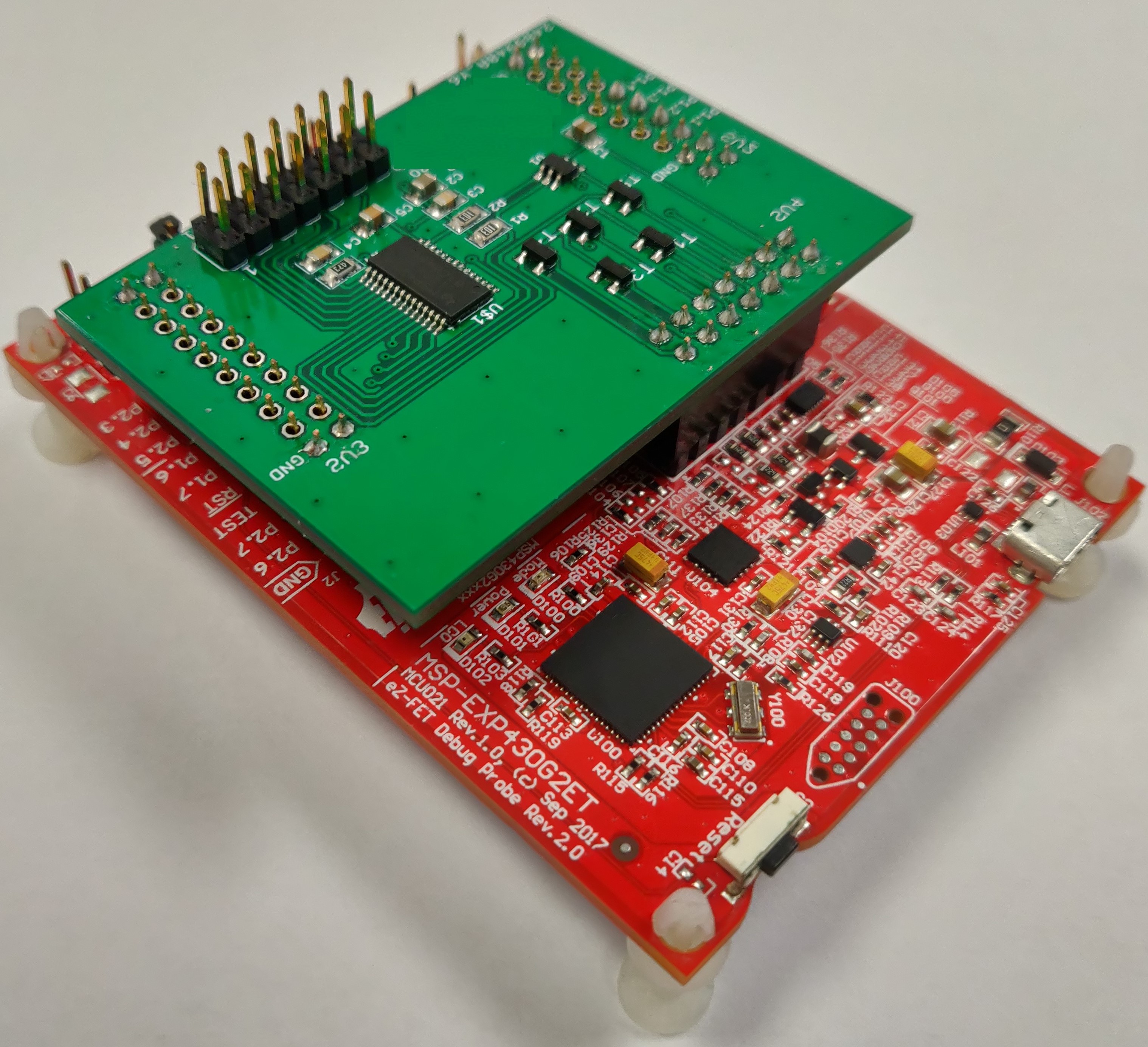}
  \caption{\footnotesize The experimental setup with aging control daughter board (top) and the MSP430G2553 Launchpad (bottom).}
  \label{fig:boards}
  \hrulefill
\end{figure}

To evaluate how effective \sys{} is at discriminating between new and recycled devices, we age a set of microcontrollers using a benchmark set of common embedded system programs.
To take advantage of voltage-accelerated aging, we evaluate \sys{} using a set of Texas Instruments MSP430G2553s~\cite{g2553datasheet}.
The MSP430G2553 is particularly suited to accelerated aging because it exposes the \textit{unregulated} supply voltage rail on a pin.
MSP430-family devices are also popular and commonly deployed in general-purpose embedded applications, making them highly representative of current commercial microcontrollers that counterfeiters repackage and sell.
We calculate the \textit{acceleration factor} as shown in Equation~\ref{eq:nbti_aging}, using the following parameters.
\begin{equation}
    \label{eq:nbti_aging}
    AF = \left(\frac{V_{str}}{V_{nom}}\right)^{\frac{\alpha}{n}} * exp\left(\frac{E_{aa}}{k} * \left(\frac{1}{T_{str}} - \frac{1}{T_{nom}}\right) * \frac{1}{n}\right)
\end{equation}
% We calculate acceleration factor for our experiments with the following parameters:
\begin{itemize}
    \itemsep0em
    \item Gate voltage exponent $\alpha = 3.5$
    \item Time exponent $n = 0.25$
    \item Apparent activation energy $E_{aa} = -0.02 eV$
    \item Boltzmann's constant $k = 8.62 * 10^{-5} eV/K$
    \item Nominal temperature $T_{nom} = 293 K$
    \item Nominal voltage $V_{nom} = 3.3 V$
    \item Stress temperature $T_{str} = 353 K$
    \item Stress voltage $V_{str} = 4.75 V$
\end{itemize}
% \hrw{Probably take this out of list format for space}
This results in an acceleration factor of 280, \ie{} one hour of operation under the stress conditions ages the SRAM in the same way as 280 hours of operation under nominal conditions.
We age each device under test for a total of 156 hours, resulting in approximately 5 years of effective age.
Because the SRAM changes most during the initial aging period, we record power-on states frequently at the beginning: at the (effective) 30 minute mark, followed by one hour, one day, one week, and one month.
After this initial period we reduce the measurement frequency to once every 4.6 months (12 real hours).

We conduct all experiments in a TestEquity 123H Thermal Chamber~\cite{tempChamber} to control ambient temperature and drive the device's supply voltage using a custom daughter board shown in Figure~\ref{fig:boards}.
A microcontroller on the daughter board controls a Digital-to-Analog Converter (DAC) capable of powering the MSP430 and transistors to isolate the MSP430 from the debugger during high-voltage aging periods and power cycles.
To reduce the effects of thermal noise and minimize device aging during measurement, we record all measurements under the nominal operating conditions with an hour thermal-stabilization period before and after the measurements.
% 12 hours = 4.6 effective months

\subsection{Evaluation Methodology}
\label{sec:eval-method}
\begin{figure}[t]
  \centering
  \includegraphics[width=\columnwidth]{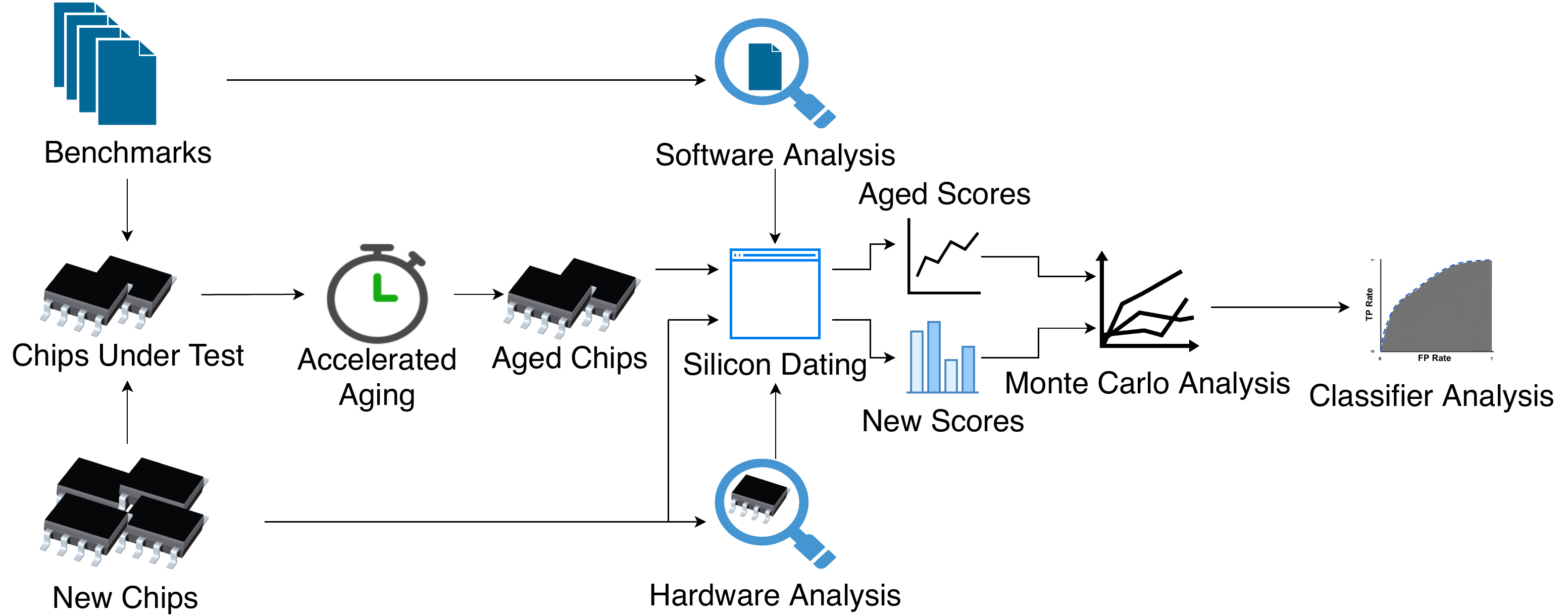}
  \caption{\footnotesize Overview of the classifier evaluation pipeline.}
  \label{fig:evaluation_pipeline}
  \hrulefill
\end{figure}
% We combine the analytical model of device aging developed in Section~\ref{sec:new_sram} with Monte Carlo methods to generalize aging results from a handful of devices to a larger test population.
% In order to generalize aging results from a handful of devices to a realistic population size, we combine the analytical model of device aging we develop in Section~\ref{sec:new_sram} with Monte Carlo~\cite{monte-carlo} numerical methods.
We combine the analytical model of device aging we develop in Section~\ref{sec:new_sram} with Monte Carlo~\cite{monte-carlo} methods to evaluate \sys{} with a large number of devices without individually aging each one.
Figure~\ref{fig:evaluation_pipeline} gives a high-level overview of our evaluation process.
We first age five devices, one running each benchmark described in Table~\ref{table:benchmarks_combined}, under accelerated-aging conditions and collect their SRAM power-on distributions after each aging interval.
% For each evaluation run of a classifier, we create a set of new-device scores by running the classifiers on the baseline set of 18 new devices used to generate the results in Table~\ref{table:new_stats} and create a normal distribution based on those scores.
To evaluate a classifier, we create a set of new-device scores by running the classifiers on the baseline set of 18 new devices used to generate the results in Table~\ref{table:new_stats}.
According to the Shapiro-Wilk~\cite{shapiro-wilk} test with $p$ = 0.05, the new-device scores for each classifier are normally distributed, allowing us to generate arbitrarily many virtual devices using the distribution from the 18 physical devices.
Finally, we apply the incremental change in score observed in the real devices at each time step to the new-device scores to match the aging progression we observe in the real devices and its effect on the classification score---Figure~\ref{fig:banding} illustrates this aging progression on a set of 100 devices.

% Evaluation stuff
We first find the upper limit of \sys{}'s effectiveness by applying the classifiers to the devices aged with fully-biased software from Section~\ref{sec:new_sram}.
We then evaluate \sys{} in stages of increasing specificity:
\begin{enumerate}
    \item In the first stage, we assume no special knowledge of the software or hardware and evaluate classification performance on devices running our five embedded benchmarks.
    \item We then evaluate the same devices but include hardware structural bias information as described in Section~\ref{sec:structural-awareness}, which allows \sys{} to leverage weakly biased cells during the scoring process.
    \item Finally, we add software-specific information to down-select useful bits and maximize the age information available to \sys{}.
\end{enumerate}
We use this evaluation process to answer several key questions about \sys{}:
\begin{enumerate}
    \item How well can \sys{} distinguish between new and recycled devices after a typical aging period?
    \item What is the minimum aging time required for \sys{} to outperform a random classifier?
    \item What factors affect \sys{}'s accuracy?
    \item What is the cost for adversaries to evade \sys{}?
\end{enumerate}

\subsection{Bounding Software-directed Aging}
\label{sec:bounding-aging}

\begin{figure}[t]
  \includegraphics[width=\columnwidth]{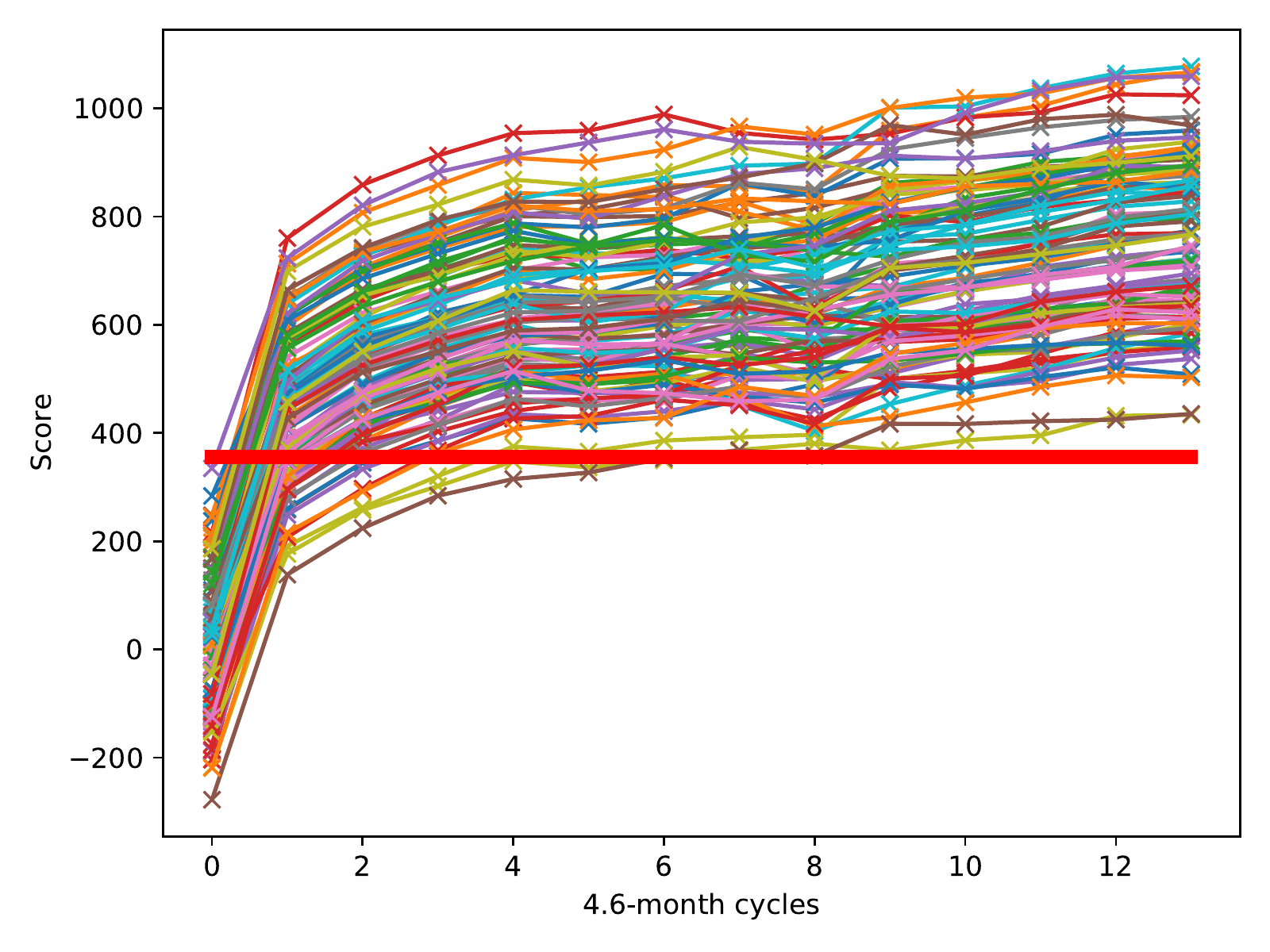}
  \caption{\footnotesize Score change over time of devices aged with fully-biased software. Horizontal line indicates the cutoff for a perfect classification decision after 5 years of aging.}
    % Source: Wide baseline data from fully_biased_scores.py -> banding.py
  \label{fig:banding}
  \hrulefill
\end{figure}

To determine \sys{}'s best case performance, we analyze the devices aged with fully-biased software in Section~\ref{sec:sram_aging}.
The difference in score between devices aged with the same software is minimal, reflecting our observation in Section~\ref{sec:new_sram} that devices respond similarly to aging.
We model the response of 1000 devices to aging using the analytical method described in Section~\ref{sec:eval-method} by applying the incremental change in score observed in the real aged devices up to the 5 year mark (the exact change modeled is a random sample from distribution of observed aging-induced changes at that time step).
This effectively tests \sys{} with 2000 devices, half of which are new and half of which are recycled.
% Source: compute relative standard deviation of output of fully_biased_scores.py

Figure~\ref{fig:banding} shows the change in score for the devices over time; we only plot 100 here for illustrative purposes.
The thicker horizontal line indicates the optimal cutoff $T$ for distinguishing between new and recycled devices after 5 years of use.
Devices with a score above this line are labeled as recycled.
For these devices \sys{} achieves a 96\% accurate classification rate after the first 4.6 month aging period, which increases to 100\% after the first 1.9 years of aging.
We explore the accuracy for shorter aging periods in Section~\ref{sec:operating-time}.
Given that electronic devices often have lifetimes on the order of decades, this suggests that \sys{} readily filters out counterfeit devices---in the extreme case.

\subsection{Figures of Merit}

We use several metrics to evaluate \sys{}'s binary classifiers when a perfect decision threshold does not exist.
\sys{} classifies devices based on the difference between the score of a device under test and the mean expected score of new devices by defining a maximum difference $T$ between the device score and the expected new score.
If $score_{dev} - score_{new} \geq T$, the device is marked as recycled.
$T$ determines the diagnostic ability of the test: a high $T$ tends towards marking more devices as new (potentially increasing false negative rate) while a low $T$ tends towards marking more devices as recycled (potentially increasing false positive rate).

To measure classifier ability we plot the Receiver Operating Characteristic (ROC) curve~\cite{intro-roc-analysis}, the True Positive Rate (TPR) vs the False Positive Rate (FPR) as $T$ varies, for each classifier under various conditions.
We vary the value of $T$ for each classifier to the point where every device tested is marked as recycled (100\% false positive rate) and where every device is marked as new (100\% false negative rate).
For reference we also plot the ROC curve for a random classifier, for which TPR and FPR always increase at the same rate as $T$ decreases.
The Area Under the ROC Curve (AUROC) enables numerical comparison of classifiers and represents the probability that the classifier will score a positive (recycled) device higher than a negative (new) one; 1.0 corresponds to a perfect classifier, while 0.5 is the AUROC for a random classifier.

We consider the ``best'' $T$ value to be the one that maximizes Informedness~\cite{informedness} as calculated in Equation~\ref{eq:informedness}  and consider the classifier accuracy at that point.
\begin{equation}
    \label{eq:informedness}
    Informedness = \frac{TP}{TP + FN} + \frac{TN}{TN + FP} - 1
\end{equation}
This choice of $T$ reflects one implicit assumption: the cost of a false positive is equal to the cost of a false negative (\ie{} wrongly rejecting a new device is as ``bad'' as wrongly accepting a recycled one).
In practice, false positives (returning a good IC) are often highly preferred to false negatives (train brake controller fails).
The optimal $T$ also depends on the recycled device prevalence, where a stricter $T$ detects more recycled devices than it rejects new ones if the test batch includes a large proportion of recycled devices.
Thus $T$ is application specific.

\subsection{Software-unaware Dating}
\label{sec:unaware-eval}
\begin{figure}[t]
  \includegraphics[width=\columnwidth]{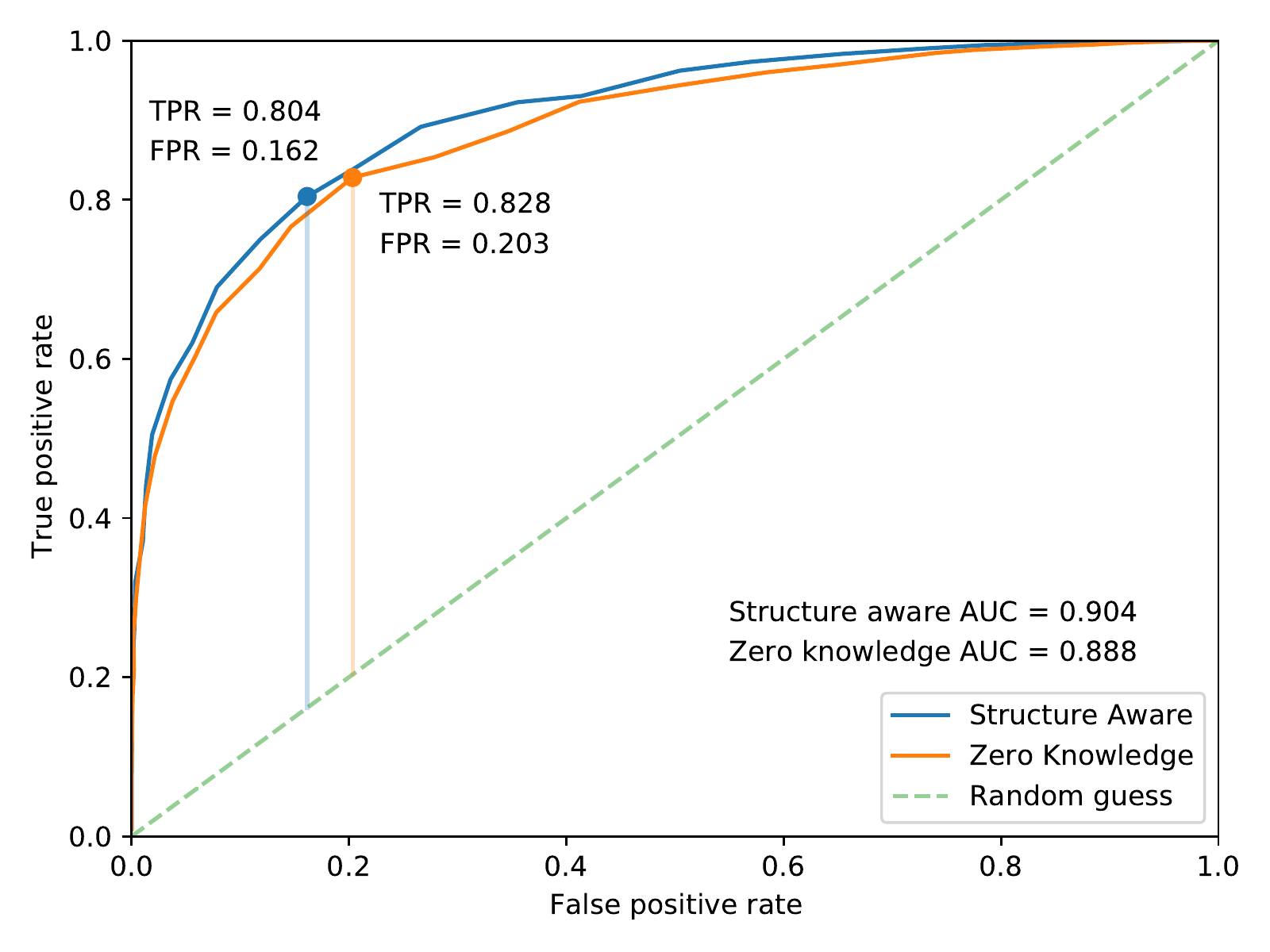}
  \caption{\footnotesize Zero-knowledge and structure-aware classifier ROC curves for a batch containing devices aged with all benchmarks.}
  \label{fig:unaware_rocs}
  \hrulefill
\end{figure}
% Source: data from all_boards_benchmarks_two_thresholds.py plotted in roc_curves.py
\begin{table}[t]
\footnotesize
\begin{tabular}{l | c c c | c c c}
                % & \multicolumn{3}{c}{Structure-unaware} & \multicolumn{3}{c}{Structure-aware} \\	
                & \multicolumn{3}{c}{Zero-knowledge} & \multicolumn{3}{c}{Structure-aware} \\	
\textbf{Benchmark}       & AUROC & TPR & Acc.    & AUROC & TPR & Acc.\\
\hline
\hline
FFT             & 0.883 & 0.798 & 80.8\%            & 0.879 & 0.793 & 80.6\%    \\
FSM             & 0.987 & 0.954 & 94.4\%            & 0.991 & 0.962 & 95.3\%     \\
FIR             & 0.777 & 0.701 & 72.5\%            & 0.854 & 0.835 & 79.1\%    \\
Quicksort       & 0.964 & 0.886 & 90.6\%            & 0.955 & 0.853 & 89.0\%    \\
PID             & 0.850 & 0.770 & 77.0\%            & 0.845 & 0.815 & 76.5\%    \\
\hline
\textbf{Mean}   & 0.892 & 0.822 & 79.3\%            & 0.905 & 0.852 & 84.1\%    \\
\end{tabular}
\caption{\footnotesize Zero-knowledge and structure-aware classifier performance.}
% Accuracy can be derived from TPR and FPR and I think the TPR/FPR numbers may "look" better
% Maybe show TPR and TNR: Good high numbers
% Source: many_rocs.py
\label{table:unaware_results}
\hrulefill
\end{table}

% \begin{table}
% \footnotesize
% \begin{tabular}{l | c c c c | c c c c}
%                 & \multicolumn{4}{c}{Structure-unaware} & \multicolumn{4}{c}{Structure-aware} \\	
% \textbf{Benchmark}       & AUROC & TPR & FPR & Acc.    & AUROC & TPR & FPR & Acc.\\
% \hline
% \hline
% FFT             & 0.880 & 0.916 & 0.316 & 80.0\%            & 0.870 & 0.813 & 0.195 & 80.9\%    \\
% FSM             & 0.985 & 0.938 & 0.043 & 94.8\%            & 0.989 & 0.949 & 0.039 & 95.5\%     \\
% FIR             & 0.775 & 0.611 & 0.202 & 70.5\%            & 0.854 & 0.775 & 0.212 & 78.1\%    \\
% Quicksort       & 0.964 & 0.883 & 0.072 & 90.5\%            & 0.957 & 0.896 & 0.100 & 89.8\%    \\
% PID             & 0.848 & 0.849 & 0.320 & 76.4\%            & 0.844 & 0.743 & 0.219 & 76.2\%    \\
% String          & 0.680 & 0.635 & 0.414 & 63.5\%            & 0.754 & 0.628 & 0.239 & 69.5\%    \\
% \hline
% \textbf{Mean}   & 0.855 & 0.805 & 0.228 & 79.3\%            & 0.878 & 0.801 & 0.167 & 81.7\%    \\
% \end{tabular}
% \caption{\footnotesize Software-unaware classifier performance with and without structural information.}
% % Accuracy can be derived from TPR and FPR and I think the TPR/FPR numbers may "look" better
% % Maybe show TPR and TNR: Good high numbers
% % Source: many_rocs.py
% \label{table:unaware_results}
% \hrulefill
% \end{table}
% We evaluate \sys{} for a large set of devices (2,000 per benchmark, half new and half aged) using the analytical evaluation framework described in Section~\ref{sec:eval-method}.
We evaluate \sys{} for a large set of devices (1,000 new and 1,000 aged) using the analytical evaluation framework described in Section~\ref{sec:eval-method}.
Figure~\ref{fig:unaware_rocs} shows the ROC curves for the software-unaware classifier.
To evaluate the effect of structural bias awareness as described in Section~\ref{sec:structural-awareness}, we plot the ROC curves for software-unaware classification with and without structural information.
The dot on each line represents the optimal operating point using Equation~\ref{eq:informedness}.
The per-benchmark results in Table~\ref{table:unaware_results} show that the software-unaware implementation of \sys{} is a reliable coarse-grain indicator of device age, correctly classifying on average 79.3\% of devices under test.
Leveraging the spatial autocorrelation of cell biases increases \sys{}'s accuracy to 84.1\% by including weakly-biased cells in the classification.

While individual software dictates the degree of detectable aging and therefore \sys{}'s overall performance, the variety of different software in a test batch also impacts the software-unaware classifier.
Because each program produces a different degree of burn-in to the hardware, the optimal decision cutoff $T$ is software-dependent.
The best $T$ for a set of programs may not be the best $T$ for any individual program.
For the software-unaware classifier, we choose the best $T$ on average as opposed to overfitting to any specific benchmark.
This results in a minor decrease in accuracy---the structure-aware classifier's ``batch accuracy'' considering all benchmarks at once is 82.1\%, compared to the 84.1\% average for individual benchmarks---but does not affect the AUROC.

\subsection{Software-aware Dating}
\label{sec:aware-eval}
\begin{figure}[t]
  \includegraphics[width=\columnwidth]{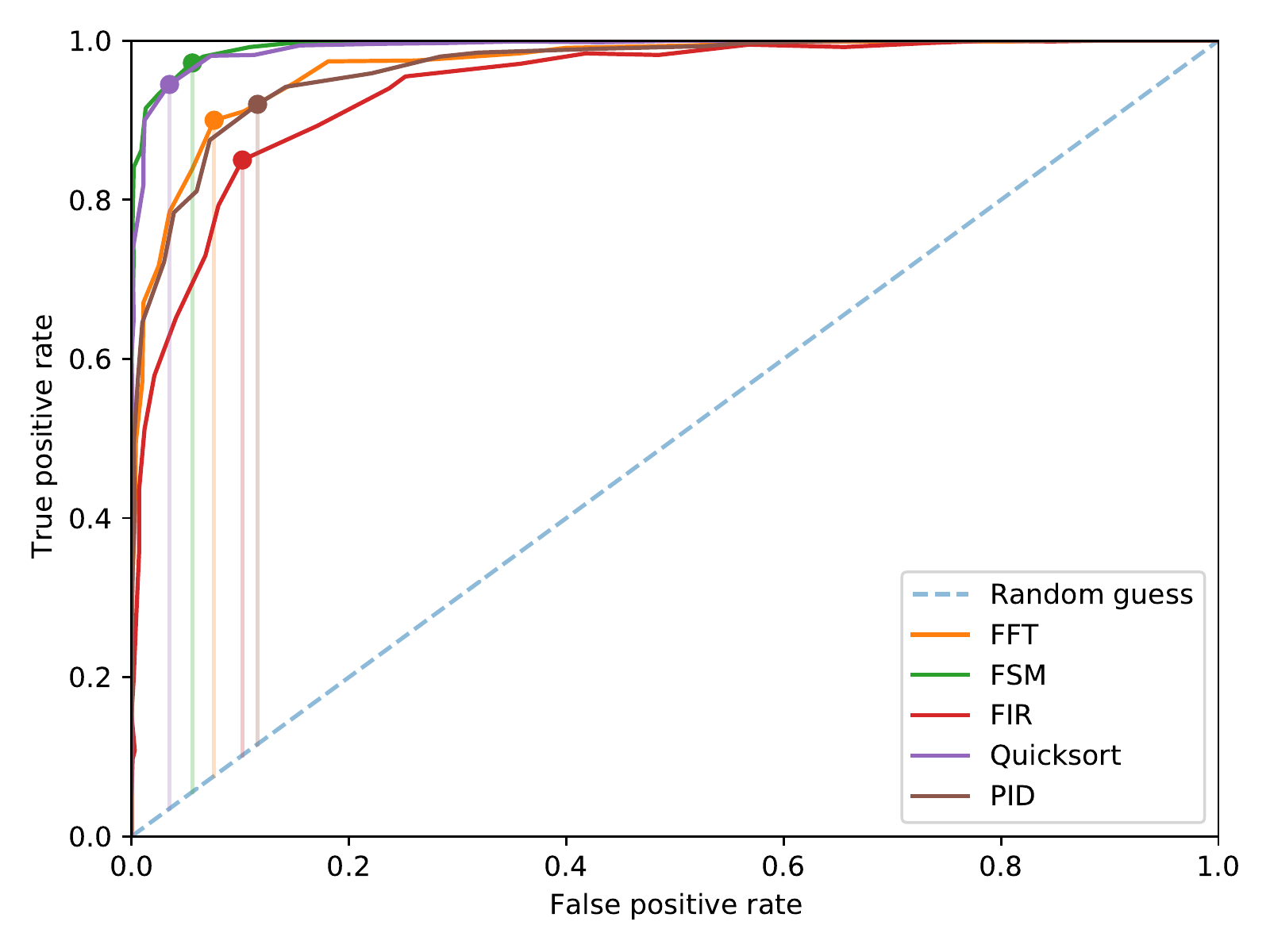}
  \caption{\footnotesize Software-aware classifier ROC curves for each benchmark.}
  \label{fig:aware_rocs}
  \hrulefill
\end{figure}
% Source: data from all_boards_benchmarks_two_thresholds.py plotted in many_rocs.py
\begin{table}
\footnotesize
\begin{tabular}{l | c c c | c c c}
                & \multicolumn{3}{c}{Structure-unaware} & \multicolumn{3}{c}{Structure-aware} \\	
\textbf{Benchmark}       & AUROC & TPR & Acc.    & AUROC & TPR & Acc.\\
\hline
\hline
FFT             & 0.968 & 0.900 & 91.2\%            & 0.974 & 0.925 & 92.4\%    \\
FSM             & 0.994 & 0.972 & 95.8\%            & 0.994 & 0.972 & 95.8\%     \\
FIR             & 0.941 & 0.850 & 87.4\%            & 0.944 & 0.858 & 87.8\%    \\
Quicksort       & 0.991 & 0.955 & 95.5\%            & 0.991 & 0.954 & 95.4\%    \\
PID             & 0.965 & 0.920 & 90.2\%            & 0.954 & 0.889 & 88.9\%    \\
\hline
\textbf{Mean}   & 0.972 & 0.919 & 92.0\%            & 0.971 & 0.920 & 92.1\%    \\
\end{tabular}
\caption{\footnotesize Software-aware classifier performance with and without structural information.}
% Accuracy can be derived from TPR and FPR and I think the TPR/FPR numbers may "look" better
% Maybe show TPR and TNR: Good high numbers
% Source: many_rocs.py
\label{table:aware_results}
\hrulefill
\end{table}

% \begin{table}
% \footnotesize
% \begin{tabular}{l | c c c c | c c c c}
%                 & \multicolumn{4}{c}{Structure-unaware} & \multicolumn{4}{c}{Structure-aware} \\	
% \textbf{Benchmark}       & AUROC & TPR & FPR & Acc.    & AUROC & TPR & FPR & Acc.\\
% \hline
% \hline
% FFT             & 0.880 & 0.916 & 0.316 & 80.0\%            & 0.870 & 0.813 & 0.195 & 80.9\%    \\
% FSM             & 0.985 & 0.938 & 0.043 & 94.8\%            & 0.989 & 0.949 & 0.039 & 95.5\%     \\
% FIR             & 0.775 & 0.611 & 0.202 & 70.5\%            & 0.854 & 0.775 & 0.212 & 78.1\%    \\
% Quicksort       & 0.964 & 0.883 & 0.072 & 90.5\%            & 0.957 & 0.896 & 0.100 & 89.8\%    \\
% PID             & 0.848 & 0.849 & 0.320 & 76.4\%            & 0.844 & 0.743 & 0.219 & 76.2\%    \\
% String          & 0.680 & 0.635 & 0.414 & 63.5\%            & 0.754 & 0.628 & 0.239 & 69.5\%    \\
% \hline
% \textbf{Mean}   & 0.855 & 0.805 & 0.228 & 79.3\%            & 0.878 & 0.801 & 0.167 & 81.7\%    \\
% \end{tabular}
% \caption{\footnotesize Software-unaware classifier performance with and without structural information.}
% % Accuracy can be derived from TPR and FPR and I think the TPR/FPR numbers may "look" better
% % Maybe show TPR and TNR: Good high numbers
% % Source: many_rocs.py
% \label{table:unaware_results}
% \hrulefill
% \end{table}
\begin{table}[t]
\centering
\footnotesize
\begin{tabular}{l c c c}
\textbf{Benchmark} & \textbf{SW 0} & \textbf{SW 1} & \textbf{Total}\\
\hline
\hline
FFT                 & 20.0\% & 1.4\% & 21.4\%\\
FSM Controller      & 73.0\% & 0.6\% & 73.6\%\\
FIR Filter          & 22.7\% & 1.1\% & 23.8\%\\
Quicksort           & 55.1\% & 1.5\% & 56.6\%\\
PID Controller      & 26.9\% & 0.9\% & 27.8\%\\
\textbf{Average}    & 39.5\% & 1.1\% & 40.6\%\\
\end{tabular}
\caption{\footnotesize Portion of software bits usable by \sys{}'s software-aware classification (where bits are ``strong'' if they are always the same value).}
% Source: benchmark_flips.py
\label{table:strong_table}
\hrulefill
\end{table}

We evaluate the software-aware implementation of \sys{} under the same conditions as the software-unaware version.
% We also fine-tune the software-aware classifier's performance by varying several parameters that are constant for the software-unaware version.\mdh{I have no clue what you are talking about here.  Too abstract. If you are trying to lead into the ensuing paragraph, this does a poor job of that.}
We also fine-tune the software-aware classifier's performance by varying the bias strength threshold required to consider a software bit usable, and setting a benchmark-specific score threshold for recycled device classification.

\paragraph{Software bit bias thresholds}

Different software bits age their associated cells at rates depending on their bias strength.
The software-aware classifier defines a minimum bias strength threshold and ignores cells associated with bits below that threshold.
Including weaker bits increases the number of cells \sys{} examines, but also decreases the chance that each bit produces a visible change in cell bias.
We evaluate the effect of different software bias thresholds by varying the minimum bit strength required for the classifier to factor the associated SRAM cell into the scoring decision.
% Our results indicate that increasing the minimum strength (based on the bit-strength formula in Section~\ref{sec:software_asymmetry}) to 1 and only accepting the strongest software bits yields an improvement over accepting all software bits: the average AUROC and accuracy when accepting all bits are 0.942 and 88.9\% respectively, which increase to 0.972 and 92.0\% with the strictest software bit threshold.
Our results indicate that increasing the minimum strength based on the bit-strength formula in Section~\ref{sec:software_asymmetry} from 0 (accepting all software bits) to 1 (accepting only the most-biased software bits) increases classification performance: the average AUROC and accuracy when accepting all bits are 0.942 and 88.9\%, respectively, which increase to 0.972 and 92.0\%, respectively, with the strictest software bit threshold.
We evaluate \sys{}'s software-aware classifier using this strict bit threshold.

\paragraph{Decision Thresholds}

The score change expected in an aged device is benchmark-dependent: generally, software with more strongly biased bits produces a larger score change in aged devices when compared to new ones.
The number of bits under consideration also affects the score variation among new devices---introducing more cells introduces more variation in the new-device population.
Benchmark-specific threshold values affect the classifier's accuracy, but not the ROC curve: switching from a single threshold value to benchmark-specific ones increases the software-aware classifier's accuracy from 91.2\% to 92.0\%.

Figure~\ref{fig:aware_rocs} and Table~\ref{table:aware_results} illustrate our performance evaluation for the software-aware classifier using the optimal parameters discussed above.
We make two observations about software-aware classification based on these results.

First, software information significantly increases the classification performance of \sys{} by maximizing the information available from each cell.
The software-aware classifier ignores weak bits and better leverages strong-$1$ bits compared to the software-unaware classifier's, which assumes only strong-$0$ bits.
Table~\ref{table:strong_table} illustrates the proportion of SRAM containing usable bits in each benchmark using the strictest possible software bit threshold.
Because these strong bits produce the most significant cell bias changes, \sys{} is better at detecting devices running programs with more usable bits (FSM) compared to those with fewer (FIR).
The count of usable bits in a program serves as a rough predictor of \sys{}'s performance.

Second, structural information has little impact on software-aware classification performance.
Table~\ref{table:aware_results} shows that combining structure-awareness and software-awareness has similar performance to having software-awareness alone.
These results imply that unknown software bit bias, rather than SRAM cell aging visibility, is the dominant noise source for \sys{}.

\subsection{Unbiased Software}
\label{sec:unbiased-software}

% \begin{figure}
% \centering
% \subfloat[Bias heatmap]
% {
%     \includegraphics[width=.4\columnwidth]{Figures/sim_biases/sim_bias_string10000.pdf}
%     \label{fig:string_bias}
% }
% \subfloat[ROC curves]
% {
%     \includegraphics[width=.5\columnwidth]{Figures/string_rocs.pdf}
%     \label{fig:string_roc}
% }
% \caption{\footnotesize Bias heatmap and ROC curves for the string processing benchmark.}
% \label{fig:string_stuff}
% \hrulefill
% \end{figure}

\begin{figure}[t]
  \includegraphics[width=\columnwidth]{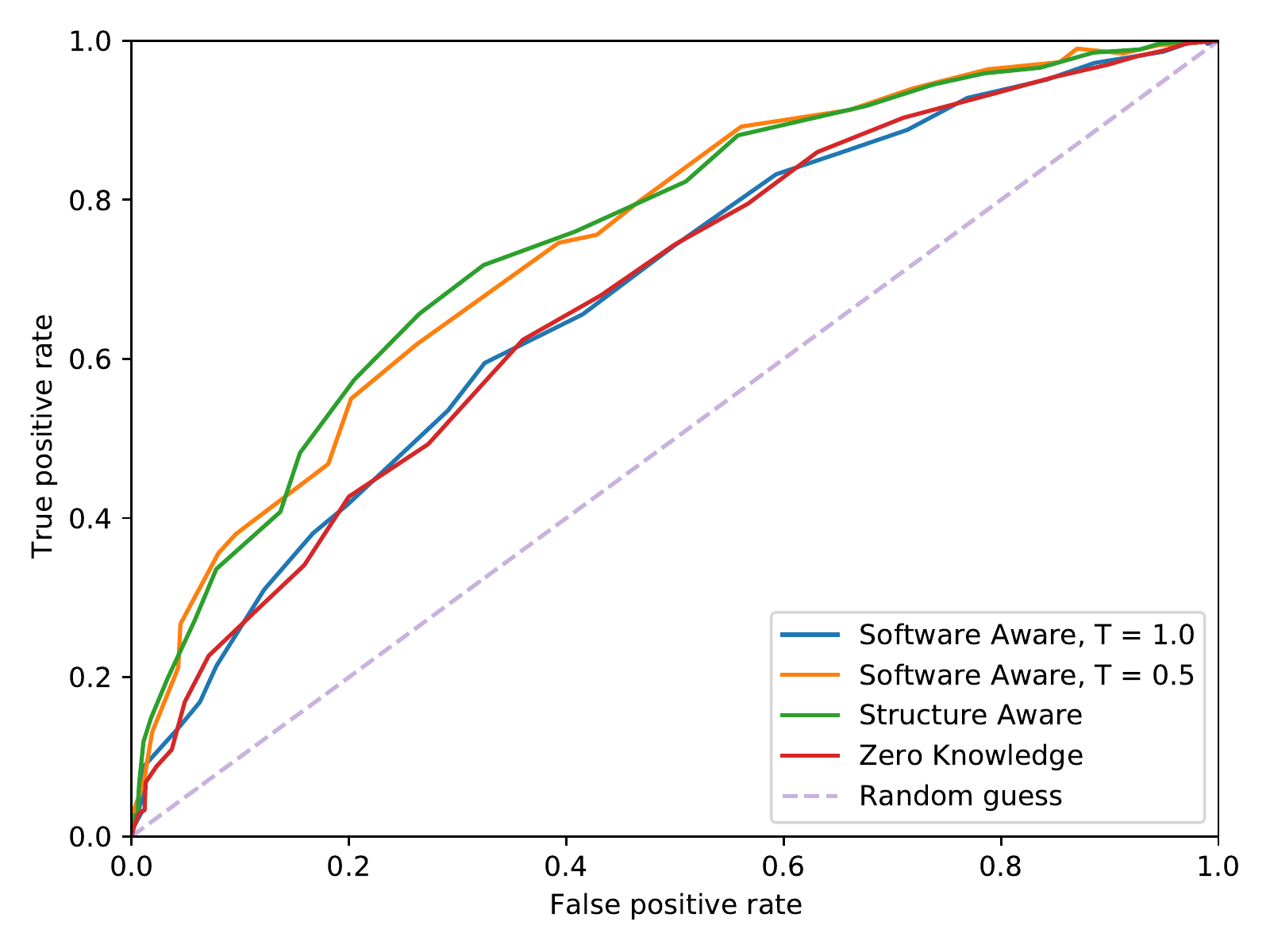}
  \caption{\footnotesize Classification performance on the text benchmark.}
  \label{fig:string_roc}
  \hrulefill
\end{figure}
% Source: data from all_boards_benchmarks_two_thresholds.py plotted in roc_curves.py

\sys{} leverages a fundamental assumption about software: software is more likely to write $0$s to SRAM than $1$s.\footnote{In actuality, \sys{} relies on bit value asymmetry (e.g., more $1$s than $0$s), it does not matter what the dominant bit value is.}
We seek to explore the impact on \sys{} when software violates this assumption.
While Section~\ref{sec:software_asymmetry} shows that the assumption does hold for many software exemplars, we identify two examples of programs that violate this assumption: (1) text-based programs that employ a compact encoding scheme and (2) cryptographic programs.
We posit that devices aged running software where a significant portion of the workload is handling/creating this unbiased data are more difficult to detect for \sys{}, because of two reasons: (1) fewer SRAM cells contain software bits biased enough to change the power-on state and (2) the software values are symmetric in terms of the number of $0$s and $1$s and will age the SRAM symmetrically.
% Even a checker board program with alternating 0's and 1's will pose a problem, as the maximum score is the same as the new device score...50/50

To evaluate the impact of unbiased software on \sys{}, we evaluate an extreme case using a text-based benchmark that fills the SRAM with ASCII (American Standard Code for Information Interchange) characters based on their relative frequency in English text.
Compared to our five benchmarks, the text benchmark has the fewest (13.1\%) usable bits and the mean bias closest to 50\% at 35\%.
\sys{}'s classification performance reflects these properties, shown in Figure~\ref{fig:string_roc}.
The structural-aware classifier achieves 69.7\% accuracy, while the zero-knowledge and software-aware classifier both provide 63.6\% accuracy.

Because most bits in the text benchmark are close to evenly biased, we also evaluate the software-aware classifier with the bias strength threshold as calculated in Section~\ref{sec:software_asymmetry} reduced from 100\% (analyzing only bits that always contain a single value) to 0\% (analyzing all software bits).
In this case, reducing the strength threshold improves performance: the low-threshold software-aware classifier is 68.9\% accurate with a ROC curve comparable to the structural-aware version.
Given the under-performance of the strict-threshold software-aware implementation compared to the software-unaware versions, we observe that the software-aware classifier is superior to the software-unaware versions when there are a sufficient number of usably biased bits in the software under test.

\subsection{Effect of Operating Time}
\label{sec:operating-time}

\begin{figure}[t]
  \includegraphics[width=\columnwidth]{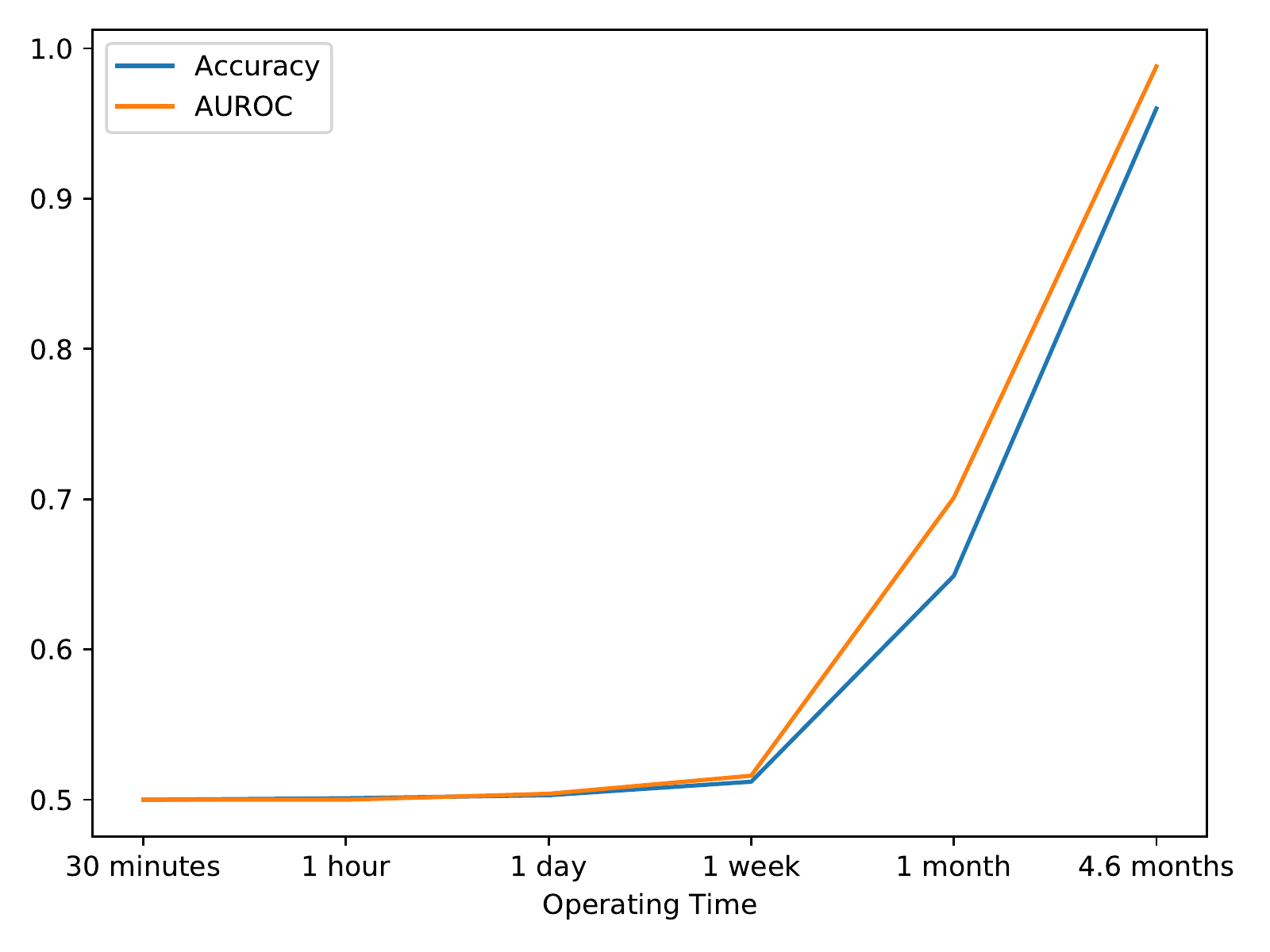}
  \caption{\footnotesize Fully-biased classification performance for short aging periods.}
  \label{fig:vs_time}
  \hrulefill
\end{figure}
% Source: data from all_boards_short_aging.py evaluated in roc_vs_time.py and plotted in vs_time.py

Several additional factors beyond the behavior of \sys{} itself affect the performance of both classifiers.
% The first is operating time: older devices are easier to detect than younger ones, as SRAM has longer to take on software's asymmetric data patterns.
The first is operating time: older devices are easier to detect than younger ones, as software has longer to burn into SRAM.
While increasing aging time makes the classification decision easier, we find that the majority of detectable aging happens early in the device's life: over 60\% of the total change in score over the 5 year aging period occurs in the first 4.6 months.
% Source: sum(score@cycle13 - score@cycle1) / 6 with unaware_narrow scores
After the first 4.6-month aging period the average AUROC and accuracy for the software-aware classifier are 0.897 and 82.8\% respectively, and 0.796 and 72.9\% for the software-unaware version.
% Source: Unaware: output of roc_curves.py with roc_data_unaware_cycle2.txt input
% Source: Aware: Average of outputs of many_rocs.py with roc_data_aware_cycle2.txt input
In the following aging periods, the AUROC and accuracy for both versions slowly increase.
% Although we continue the aging experiments for the full five years, \sys{} achieves a perfect classification rate for fully-biased devices after the first nine months.

To determine the minimum aging time required for \sys{} to detect a recycled device, we evaluate the devices running fully-biased software at each of the short aging times leading up to the first 4.6 months described in Section~\ref{sec:setup}.
Figure~\ref{fig:vs_time} illustrates the classification performance as a function of time; performance begins to increase after the first week of use, reaching 64.9\% after the first month of aging.
Based on the data shown in Figure~\ref{fig:vs_time}, we estimate 1 month to be the minimum operating time for \sys{} to achieve better-than-random detection rates for devices running realistic software.
While \sys{} is less accurate for ``younger'' devices, we note that such devices are also significantly less likely to fail prematurely than the older devices \sys{} reliably detects.

% What's the minimum operating time for detection? If a device operated for a week it's not even really that worn out. Our detector is better at detecting severely aged devices, which is what we care about.

\subsection{Effect of New-device Variance}
\label{sec:baseline-variance}

\begin{table}[t]
\footnotesize
\begin{tabular}{l | c c c}
\textbf{Baseline}  & Original & Low Variation & High Variation \\
\hline
Mean bias $\sigma{}$*       & 0.015     & 0.007 & 0.0304 \\
Portion strong 1 $\sigma{}$ & 1.2\%     & 0.76\% & 3.0\% \\
Portion strong 0 $\sigma{}$ & 1.9\%     & 0.58\% & 3.3\% \\
\hline
Zero-knowledge AUROC        & 0.892 & 0.990 & 0.738    \\
Zero-knowledge accuracy     & 83.1\% & 96.7\% & 67.9\%   \\
\hline
Structure-aware AUROC            & 0.905     & 0.994 & 0.746 \\
Structure-aware accuracy         & 84.1\%    & 97.3\% & 68.4\% \\
\hline
SW-aware AUROC              & 0.972     & 0.999 & 0.875 \\
SW-aware accuracy           & 92.0\%    & 98.9\% & 80.4\% \\
\end{tabular}
\caption{\footnotesize Summary statistics and classifier performance for high and low deviation baselines. * Standard deviation}
\label{table:deviation}
\hrulefill
\end{table}

\sys{}'s performance depends on software behavior and the degree of SRAM burn-in, but also on the degree of variability in the power-on SRAM state statistics of new devices.
Because aging affects each device similarly---the change in age score is independent of a device's bias when it was new---\sys{} performs worse for device families with high inter-device variance in SRAM power-on states (which corresponds to higher variance in their age score).
We measure the impact this has on \sys{}'s classification ability by down-selecting devices from the new device population to form higher- and lower-variation sets and use those sets as baselines to re-evaluate \sys{}.

Table~\ref{table:deviation} details the changes in summary statistic variation as well as \sys{}'s performance using each baseline set.
The low-variance baseline predicts \sys{}'s classification performance for device families such as the MSP432, which has lower variation between new devices (Table~\ref{table:new_stats}) than the set of MSP430s chosen for low variance.
The high-variation baseline evaluation highlights the advantage of using software analysis to down-select bits in aging classification: a baseline set that approximately doubles the standard deviations of the summary statistics in Table~\ref{table:deviation} reduces both the zero-knowledge and structure-aware classifiers' AUROC and accuracy by approximately 0.155 and 15.5\%, respectively.
The software-aware classifier is more resilient to variance---switching to the high-variation baseline reduces the software-aware classifier's AUROC and accuracy respectively by 0.097 and 11.6\%, roughly 2/3 the performance loss of the software-unaware classifiers.
These results show that while both implementations of \sys{} work for high-variance baseline populations, software-awareness further reduces its impact and makes \sys{} applicable to a wider variety of platforms.

\subsection{Effect of Natural V\textsubscript{th} Recovery}
\label{sec:natural_recovery}
Transistors subjected to NBTI aging partially recover (\ie{} their V\textsubscript{th} falls) when the transistor is not conducting, but most degradation is permanent~\cite{nbti-recovery}.
V\textsubscript{th} recovery is reflected in SRAM power-on states as a return to the pre-age power-on bias dictated by hardware variation instead of software bias, reducing the measurable difference between recycled and new devices and makes the classification decision harder.
The most common scenario where transistor recovery affects a device to a significant degree is between uses: for example, a recovered device is sold to a third-party broker where it sits unused for a period of time before a customer buys it.

We measure the effect recovery has on \sys{} by leaving the benchmark-aged devices unpowered at room temperature for six weeks after the last aging period, then measuring the power-on states and applying our classifiers to the new data.
Six weeks encompasses the majority of device recovery, as recovery follows a logarithmic trend (i.e., almost no additional recovery after 1 day)~\cite{attacking-puf}.
Our results show that recovery has a minimal effect on \sys{}, reducing accuracy by approximately 7\% for all classifiers, with smaller reductions in AUROC.
These results indicate that the SRAM aging effect measured by \sys{} acts as a lasting marker of device age.

\subsection{Adversarial V\textsubscript{th} Recovery}
\label{sec:adversarial_recovery}
While natural V\textsubscript{th} recovery has a minor effect on \sys{}, sufficiently motivated adversaries may manipulate the SRAM aging effects \sys{} measures to obscure recycled devices by aging SRAM cells back towards their original power-on state.
The counter-aging process would mirror our own accelerated-aging procedure in Section~\ref{sec:setup}, but with different bit biases to cancel out the naturally burnt-in software biases.
We identify several requirements for successful counter-aging:
\begin{itemize}
    \item \textbf{The adversary must integrate the devices into a counter-aging system}: counter-aging requires characterizing devices, developing counter-aging software, and integrating each device into hardware to flash the software and control aging. Adversaries must monitor the aging process to ensure the accelerated-counter-aging procedure does not damage the device, already worn from legitimate field use, in a way detectable by functional tests. From our own experiences on this project, this is challenging and costly to scale to 100's of devices.
    \item \textbf{The adversary must be able to accelerate aging}: counteracting years of device age in a time-efficient manner requires accelerating aging based on Equation~\ref{eq:nbti_aging} to bring the V\textsubscript{th} of the relatively unused transistors in line with the naturally aged ones.\footnote{While the naturally aged transistors experience some degree of V\textsubscript{th} recovery, the logarithmic nature of recovery means the counter-aged transistors will recover approximately the same amount shortly after counter-aging: a counterfeiter must produce similar amounts of permanent NBTI wear in each transistor.} This makes counter-aging devices with integrated voltage regulators impractical, because they clamp the voltage seen by the SRAM to a fixed level---eliminating the most powerful aging accelerant. For example, counteracting 5 years of device age at room temperature (20\degree{}C) on the MSP432 requires approximately 2.8 years of running at the maximum operating temperature of 85 \degree{}C. Even with unfettered access to SRAM's power rail, countering just the 5 years of aging in our experiments requires the adversary to spend 7 days counter-aging per device, 21 days for a more realistic deployment of 15 years.
\end{itemize}
Sufficiently motivated malicious counterfeiters with the required resources can overcome the above barriers for thwarting \sys{}, which only acts as a deterrent in these cases.
However, the majority of counterfeiters are profit-driven, not target-driven: the costly time and infrastructure requirements to counter-age devices serve as a strong disincentive to defeating \sys{}.

\section{Related Work}
\label{sec:related}

% \mdh{We could use some more citations.  This would be a good place to add a bunch.}

Given the threat counterfeit devices pose to a variety of areas, past work addresses recycled and other counterfeit IC detection using a wide range of approaches.
In this section, we divide the current body of research into four broad categories.

\subsection{Physical Tests/Inspection}
Many of the most widely deployed techniques capable of detecting counterfeit electronics are physical tests in the form of material analysis~\cite{counterfeit-mitigation, counterfeit-detection} or inspections using techniques such as Scanning Electron Microscopy (SEM)~\cite{counterfeit-integrated-circuits}.
These techniques capture a wide range of potential sources of IC failure and are often available as part of failure analysis labs~\cite{physical-test}.
However, physical testing suffers from several challenges that limit its application in counterfeit detection.
% Simple external tests such as blacktop or hermeticity testing, which are sensitive to manufacturing defects typical of counterfeiting operations, often fail to detect recycled ICs because such devices were built to the OEM's specifications.
Simple external tests such as blacktop or hermeticity testing are sensitive to manufacturing defects typical of counterfeiting operations, but fail to detect recycled ICs because such devices were built to the OEM's specifications.
More effective techniques often include destructive tests such as high temperature/humidity testing~\cite{qualification-testing} or IC delayering, which restricts the test to a small sample portion of devices and reduces the likelihood of finding a few counterfeit ICs within a larger set.
Physical testing is also normally performed by hand using expensive specialized equipment, making large batch testing impractical.

\subsection{Parametric Classification}
Electrical parameter testing solves many of the problems endemic to physical inspections because it can be automated; electrical characteristics such as maximum frequency, operating current, and minimum voltage vary enough between manufacturing processes to distinguish counterfeit from legitimate devices.
Counterfeit IC detection techniques based on electrical parameters characterize new ICs and classify devices under authentication as counterfeit if they do not fit the known-good profile~\cite{path-delay, statistical-methods, dynamic-current}.
Parametric analysis offers a low overhead, scalable solution for detecting out-of-spec or up-marked counterfeit ICs, as a device can quickly be characterized with several parametric measurements.
However, parametric techniques are ill-suited for recycled IC detection because they are highly sensitive to systematic variation: frequency, power consumption, and other electrical parameters vary significantly enough across new devices that the parameters for aged devices are not statistically distinct from new ones.
For recycled microcontrollers, the best non-destructive parametric approach has a correct classification rate of 19.2\%~\cite{statistical-methods}.

Of the different counterfeit detection strategies, \sys{} is most comparable to parametric classification techniques in terms of time and infrastructure overhead.
The key difference is that the tightly-coupled nature of SRAM cells insulates the cell bias statistics from systematic variation by treating it as common-mode noise within a cell.
The primary source of noise for \sys{} is random chaotic variation between individual transistors, which is minimal compared to the measurable change introduced by age.

\subsection{Aging Detection Circuits}
% The challenges of recycled IC detection using physical and parametric techniques have driven researchers to purpose-built solutions.
To address the challenges of recycled IC detection using physical and parametric techniques, past work also explores purpose-built aging-detection hardware.
One solution is to embed circuits into the IC that respond predictably to aging effects and compare their behavior to a known new model.
Typical circuits used for these systems include fuses and anti-fuses~\cite{p-val, on-chip-structures}, Schmitt-Triggers~\cite{schmitt-trigger}, and ring oscillators~\cite{fingerprint-sensor, fpga-aging-monitor, fpga-aging-analysis, silicon-odometer}.
Purpose-built circuits enable high-confidence detection of recycled devices even if the device under test has only been used for a few days; ring-oscillator based systems achieve $>$99\% accuracy within a month of moderate use~\cite{on-chip-structures}.
These systems typically target FPGAs because of the low cost of adding aging-detection circuitry, but can be designed in to other ICs where counterfeit recycling is expected to be a problem.
Unfortunately, we are not aware of any systems that include aging-detection hardware, especially the older, discontinued, designs at particular risk of counterfeiting.
Thus aging-detection hardware primarily targets future niche designs.

\subsection{Physical Unclonable Functions (PUFs)}
% In order to detect recycled ICs without additional hardware, researchers have turned to PUFs derived from intrinsic features unique to each device.
PUF-based solutions allow recycled IC detection without additional hardware.
Hardware-free PUFs for recycled IC detection characterize a device's SRAM by identifying aging-insensitive cells to serve as a permanent ID and aging-sensitive cells to indicate a device is recycled~\cite{scare, zero-cost-sram}, and achieve detection rates similar to hardware solutions.
However, hardware-free anti-counterfeiting PUFs ultimately suffer from the same limitations hardware-based solutions do: they do not address old devices or those built without counterfeit protection in mind, and come with the additional burden of device enrollment, tracking, and protection of the device ID.

\section{Conclusion}
\label{sec:conclusion}
Our evaluation of commercial microcontrollers running common embedded workloads highlights the predictable impact of software on SRAM's analog properties.
\sys{} exploits the combination of asymmetric software data and software-directed aging of SRAM to make a classifier that accurately discriminates between new and recycled devices.
Our experimental results show that \sys{} is effective at identifying even lightly used devices, is robust to both common noise sources, and imposes an impractically high bar for an adversary who attempts to reverse SRAM aging.

\sys{} measures the predictable consequences of widespread trends in both hardware and software design to answer the question: \textit{how can we reliably detect recycled devices that were not designed with counterfeit-prevention in mind?}
\sys{} is low-cost, as it uses easily measurable digital-domain channels to reveal analog-domain changes, and expandable, as it uses application-specific information about both the hardware and software to refine the aging model and maximize the information revealed by a device under test.
Our results motivate a new approach to broadly-applicable counterfeit detection techniques based on modeling hardware's natural change over time, driven by its typical use case.

\section*{Acknowledgements}
The project depicted is sponsored by the Defense Advanced Research Projects Agency.
The content of the information does not necessarily reflect the position or the policy of the Government, and no official endorsement should be inferred.
Approved for public release; distribution is unlimited.

%-------------------------------------------------------------------------------
\bibliographystyle{plain}
\bibliography{references}

%%%%%%%%%%%%%%%%%%%%%%%%%%%%%%%%%%%%%%%%%%%%%%%%%%%%%%%%%%%%%%%%%%%%%%%%%%%%%%%%
\end{document}